\documentclass [12pt]{article}

\usepackage{tabularx}
\usepackage[english]{babel}
\usepackage[dvips]{graphicx}
\usepackage{indentfirst}
\usepackage{epsfig}
\begin{document}

\title{
Monte Carlo Simulation of a NC Gauge Theory on The Fuzzy Sphere }

\author{Denjoe O'Connor$^{a}$, Badis Ydri$^{b}$\footnote{Current Address : Institut fur Physik, Mathematisch-Naturwissenschaftliche Fakultat I, Humboldt-universitat zu Berlin, D-12489 Berlin-Germany.}\\
$^{a}$School of Theoretical Physics, Dublin Institute for Advanced Studies\\
 Dublin, Ireland.\\
$^{b}$Department of Physics, Faculty of Science, Badji Mokhtar-Annaba University,\\
 Annaba, Algeria.}

\maketitle

\begin{abstract}
We find using Monte Carlo simulation  the phase structure of noncommutative $U(1)$ gauge theory in two dimensions with the fuzzy sphere ${\bf S}^2_N$ as a non-perturbative regulator. There are three phases of the model. $i)$ A matrix phase where the theory is essentially $SU(N)$ Yang-Mills reduced to zero dimension . $ii)$ A weak coupling fuzzy sphere phase with constant specific heat and $iii)$ A strong coupling fuzzy sphere phase with non-constant specific heat. The order parameter distinguishing the matrix phase from the sphere phase is the radius of the fuzzy sphere. The three phases meet at a triple point. We also give the theoretical one-loop and $\frac{1}{N}$ expansion predictions for the  transition lines which are in good agreement with the numerical data. A Monte Carlo measurement of the triple point is also given.
\end{abstract}
\tableofcontents
\section{Introduction}
Quantum noncommutative ( NC ) gauge theory is essentially unknown beyond one-loop \cite{szabor}. In the one-loop approximation of the quantum theory  we know for example that gauge models on the Moyal-Weyl spaces are renormalizable \cite{martin}. These models were also shown to behave in a variety of novel ways as compared with their commutative counterparts. There are potential problems with unitarity and causality when time is noncommuting, and most notably we mention the notorious UV-IR mixing phenomena which is a generic property of all quantum field theories on Moyal-Weyl spaces and on noncommutative spaces in general \cite{szabor,min}.  However a non-perturbative study of pure two dimensional noncommutative gauge theory was then performed in \cite{jun-wolf}. For scalar field theory  on the Moyal-Weyl space some interesting non-perturbative results using theoretical and Monte Carlo methods were obtained for example in \cite{monte1}. An extensive list of references on these issues can be found in \cite{szabor} and also in \cite{others1}

The fuzzy sphere ( and any fuzzy space in general ) provides a regularized field theory in the non-perturbative regime ideal for Monte-Carlo simulations. This is the point of view advocated in \cite{thesis}. See also \cite{madore,ars, iso} for quantum gravity, string theory or other different motivations. These fuzzy spaces consist in replacing continuos manifolds
by matrix algebras and as a consequence the resulting field theory will only have  a finite number of
degrees of freedom. The claim is that this method has the advantage 
-in contrast with lattice- of preserving all continous symmetries of the original action at least at the classical level. This proposal was applied to the scalar ${\phi}^4$ model in \cite{xavier}. Quantum field theory on fuzzy spaces was also studied perturbatively quite extensively. See for example \cite{perturbation,badis2,S2S2}. For some other non-perturbative ( theoretical or Monte Carlo ) treatement of these field theories see \cite{nonperturbative}.

Another motivation for using the fuzzy sphere is the following. The Moyal-Weyl NC space is an infinite dimensional matrix model and not a continuum manifold and as a consequence it should be regularized by a finite dimensional matrix model. In $2$ dimensions the most natural candidate is the fuzzy sphere $S^2_N$ which is a finite dimensional matrix model which reduces to the NC plane in some appropriate large $N$ flattening limit. This limit was investigated perturbatively  in \cite{badis2,badis1} for scalar and Yang-Mills field theories respectively. In $4-$dimensions we should instead consider Cartesian products of the fuzzy sphere $S^2_N$ \cite{S2S2}, fuzzy ${\bf CP}^2_N$ \cite{CP2} or fuzzy ${\bf S}^4$ \cite{S4}. An alternative way of regularizing gauge theories  on the Moyal-Weyl NC space is based on the matrix model formulation of the twisted Eguchi-Kawai model. See for example \cite{kawai,szabor1,last30}.

The goal of this article and others \cite{ref1,ref} is to find the phase structure ( i.e map the different regions of the phase diagram ) of noncommutative  $U(1)$ gauge theories in $2$ dimensions on the fuzzy sphere ${\bf S}^2_N$.  There are reasons to believe that the phase diagram of NC $U(N)$ models will be the same as that of their $U(1)$ counterparts thus we will only concentrate on the $U(1)$ models. Furthermore it seems that the nature of the underlying NC space is irrelevant. In other words $U(1)$ gauge models on the NC Moyal-Weyl plane ${\bf R}^2_{\theta}$ , on the fuzzy sphere ${\bf S}^2_N$ and on the NC torus  ${\bf T}^2_{\theta}$ will fall into the same universality class. Hence we consider solely the fuzzy sphere since it is the most convenient two dimensional space for numerical simulation. 

There seems to exist three different phases of  $U(1)$ gauge theory on ${\bf S}^2_N$. In the matrix phase the fuzzy sphere vacuum collapses under quantum fluctuations and there is  no underlying sphere in the continuum large $N$ limit. Rather we have a $U(N)$ YM theory dimensionally reduced to a point. In this phase the model should be described by a pure ( possibly a one-)matrix model without any spacetime or gauge theory interpretation. This phenomena was first observed in Monte Carlo simulation in \cite{nishimura} for $m=0$. In \cite{ref} it was shown that the fuzzy sphere vacuum becomes more stable as the
mass $m$ of the scalar normal component of the gauge field increases. Hence this vacuum becomes completely stable when  this normal scalar field is projected out from the model. This is what we observe in our Monte Carlo simulation in the limit $m{\longrightarrow}{\infty}$.

The principal new discovery of this paper is that the fuzzy sphere phase splits into two distinct regions corresponding to the weak and strong coupling phases of the gauge field. These are separated by a third order phase transition. This transition is consistent with that of a  one-plaquette  model \cite{gross}. Our results indicate that non-perturbative effects play a significant role than expected from the $1/N$ study of \cite{steinackers2}. In particular  these results indicate that quantum noncommutative gauge theory is essentially equivalent to ( some ) quantum commutative gauge theory not necessarily of the same rank. This prediction goes also in line with the powerful classical concept of Morita equivalence between NC and commutative gauge theories on the torus \cite{szabor,szabor1}. 

This article is organized as follows. 
In section $2$ we will describe the phase diagram of the NC $U(1)$ gauge model in $2$D. In section $3$ we will review the one-loop theory of the model. In section $4$ we will discuss our Monte Carlo results with some more detail. In section $5$ we will introduce the one-plaquette approximation of the model and then
we will give a theoretical derivation of the one-plaquette line.  We conclude in section $6$ with a summary and some general remarks. 
In the appendix we discuss ( among other things ) the measurement of order parameters and probability distribution.  
\section{Phase diagram}
The basic action is written in terms of three $N{\times}N$ matrices $X_a$ as follows
\begin{eqnarray}
S=N\bigg[-\frac{1}{4}Tr[X_a,X_b]^2+\frac{2i{\alpha}}{3}{\epsilon}_{abc}TrX_aX_bX_c\bigg]-Nm^2{\alpha}^2 TrX_a^2+\frac{Nm^2}{2c_2} Tr(X_a^2)^2.\label{model}
\end{eqnarray}
The basic parameters of the model are $\tilde{\alpha}=\alpha\sqrt{N}$ and $m$. The gauge coupling constant is $g^2=\frac{1}{\tilde{\alpha}^4}$. We will also need $\hat{\alpha}=\tilde{\alpha}\sqrt{1-\frac{2}{N}}$, $\bar{\alpha}=\tilde{\alpha}\sqrt{N}$ and $\bar{m}=\frac{m}{N}$. The one-loop critical values of $\tilde{\alpha}$ , $\hat{\alpha}$ and $\bar{\alpha}$ are $\tilde{\alpha}_{*}$,  $\hat{\alpha}_{*}$ and $\bar{\alpha}_{*}$ respectively. Clearly in the large $N$ limit $\tilde{\alpha}_{*}=\hat{\alpha}_{*}$ and   $\bar{\alpha}_{*}=\tilde{\alpha}_{*}\sqrt{N}$. For reasons which will become clear in the text the measured critical values are denoted as follows. The measurement of  $\tilde{\alpha}_{*}$ is denoted by ${\alpha}_s$. There are two physically distinct measurements of $\hat{\alpha}_{*}$ denoted by ${\alpha}_{ma}$ and ${\alpha}_{mi}$. In terms of $\tilde{\alpha}$ these are given by ${\alpha}_{ma}=\tilde{\alpha}_{ma}\sqrt{1-\frac{2}{N}}$ and ${\alpha}_{mi}=\tilde{\alpha}_{mi}\sqrt{1-\frac{2}{N}}$. There are two physically different measurements of  $\bar{\alpha}_{*}$ denoted by ${\alpha}_p=\tilde{\alpha}_{ma}\sqrt{N}$ and   $\bar{\alpha}_{s}={\alpha}_{s}\sqrt{N}$.

The phase diagram of the model (\ref{model}) is given in figure $(1)$. This is the central result of this article. In this section we will briefly explain the main properties of the different phases of the model. More detail will be given in the rest of the article.
\begin{figure}
\begin{center}
\includegraphics[width=13cm,angle=-90]{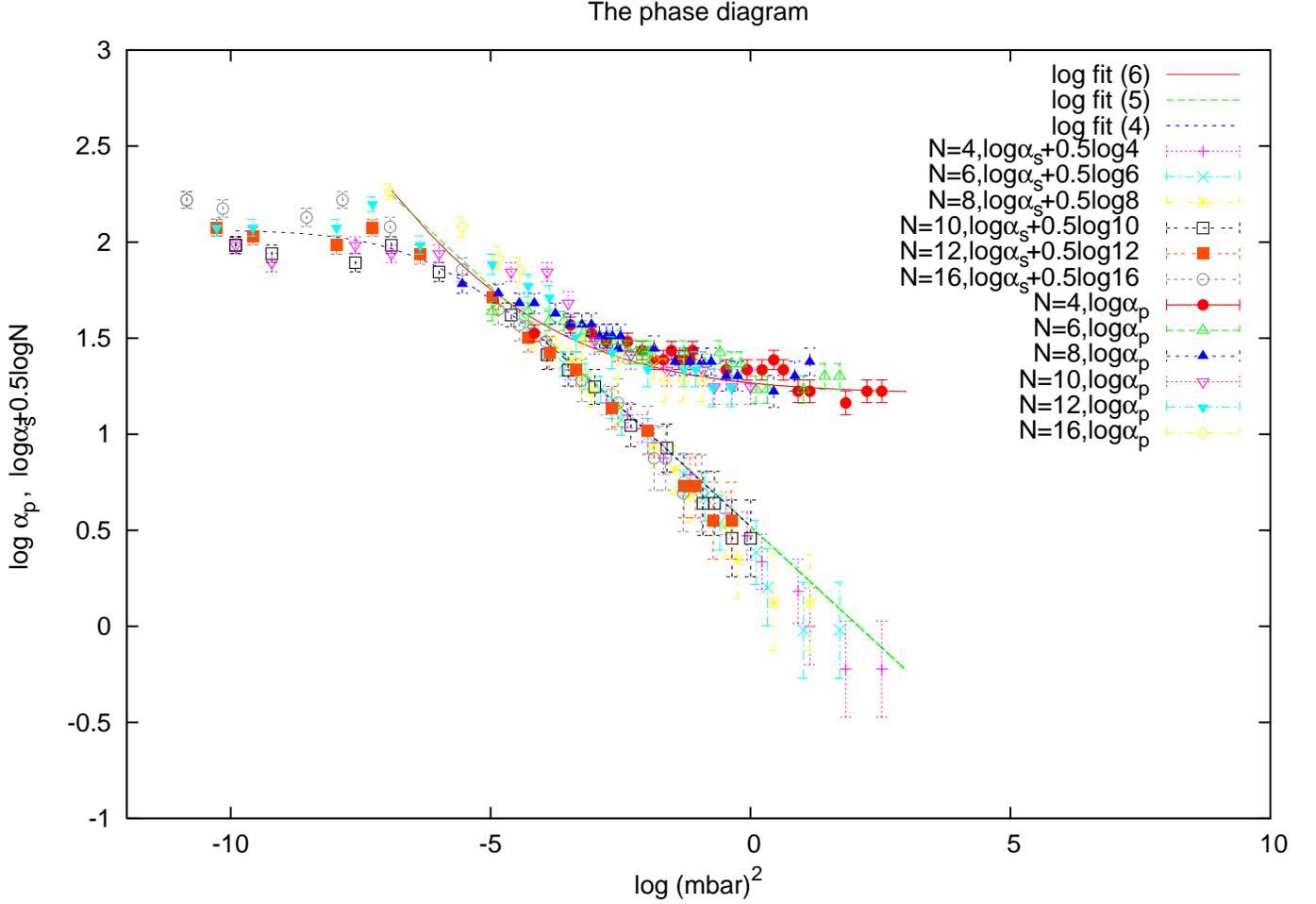}
\caption{{ The phase diagram of the model (\ref{model}). The two fits (\ref{fitintr}) and (\ref{5}) are expected to coincide very well with the data only for very large masses. The fits (\ref{5}) and (\ref{pred}) are identical for large masses ( or equivalently past the triple point ). Above the upper critical line we have a fuzzy sphere in the weak regime of the gauge theory. Between the two lines we have a fuzzy sphere in the strong regime of the gauge theory. Thus the upper critical line is the one-plaquette critical line (\ref{fitintr}). Below the lower critical line we have the matrix phase. This last line agrees very well with the one-loop prediction (\ref{5}). The one-plaquette line approaches in the limit $m{\longrightarrow}{\infty}$ a constant value given by $\log{\alpha}_p=\log(3.35)=1.21$. The triple point is also seen to exist within the estimated range. Before we reach the triple point the critical line agrees as well with the one-loop prediction (\ref{pred}). Recall that $\log\bar{m}^2=\log m^2-2\log{N}$, $\log{\alpha}_p=\log\tilde{\alpha}_{ma}+0.5\log{N}$ and $\log{\alpha}_{s}+0.5\log{N}=\log\bar{\alpha}_{s}$.}}
\end{center}\label{phase}
\end{figure}

We measure the average value of the action $<S>$  as a function of  $\tilde{\alpha}$ and we measure the specific heat $C_{v}=<S^2>-<S>^2$ as a function of  $\hat{\alpha}$ for different values of $N$. We consider $N=4,6,8,10,12,16$. In the first step of the  simulation the mass parameter $m$ is taken to be some fixed number. Then we vary the mass parameter and repeat the same experiment. The choice of $\hat{\alpha}$ for the specific heat is only due to finite size effects and has no other physical significance since in the large $N$ limit $\hat{\alpha}=\tilde{\alpha}$. 

We observe that different actions $<S>$ which correspond to different values $N$ ( for some fixed value of $m$ ) intersect at some value of the coupling constant $\tilde{\alpha}$ which we denote ${\alpha}_s$. In the limit of small masses, viz $m{\longrightarrow}0$, this intersection point marks a discontinuity in the action and it occurs around the value ${\alpha}_s=2.2$. In figure $2$ we plot the action $<S>$ versus $\tilde{\alpha}$ for $N=10,12,16$ and $m^2=0.25,3,100$. For large masses  we observe that the intersection point becomes smoother. It is also clear that the critical point  ${\alpha}_s$ decreases as we increase $m$ and  it will reach $0$ in the limit $m{\longrightarrow}\infty $. 

For the specific heat the situation is more involved. We observe in the limit $m{\longrightarrow}0$ a peak around ${\alpha}_s=2.2$ which marks a sharp discontinuity in $C_{v}$. See the first graph in figure $3$ or figure $7$. Above this critical value the specific heat is given by ${C_{v}}=N^2$ while below this critical value it is given by ${C_{v}}=0.75N^2$. The regime $\tilde{\alpha}{\geq}{\alpha}_s$ is the fuzzy sphere phase whereas  $\tilde{\alpha}{\leq}{\alpha}_s$ corresponds to the so-called matrix phase. 

As $m$ increases things get more complicated and they only simplify again when we reach large values of $m$. ${\alpha}_{ma}$ and ${\alpha}_{mi}$ are precisely the values of $\hat{\alpha}$ at the maximum and minimum values of the specific heat. Thus $\tilde{\alpha}_{ma}$ and $\tilde{\alpha}_{mi}$ are the values of $\tilde{\alpha}$ at the maximum and minimum values of the specific heat.  The peak of the specific heat moves slowly to smaller values  of $\tilde{\alpha}$ as we increase $m$. The agreement between $\tilde{\alpha}_{ma}$ and ${\alpha}_s$ for small masses is good whereas the values $\tilde{\alpha}_{mi}$ at the minimum of $C_{v}$ for small masses are significantly different from ${\alpha}_s$. Thus in this regime of small masses $\tilde{\alpha}_{ma}$ is still detecting the ${\bf S}^2_N-$to-matrix phase transition. Similarly to the case $m=0$ the specific heat  ${C}_{v}$ as a function of $\tilde{\alpha}$  is equal to $N^2$ in the fuzzy sphere phase for values of $\tilde{\alpha}$ such that $\tilde{\alpha}{\geq}\tilde{\alpha}_{ma}$ .

The physics is drastically different for large masses since the roles of $\tilde{\alpha}_{ma}$ and  $\tilde{\alpha}_{mi}$ are completely reversed. There is a shallow valley in the specific heat starting to appear for values of $\tilde{\alpha}$ inside the matrix phase as $m$ slowly increases. Furthermore as $m$ keeps increasing we observe that the peak flattens slowly and disappears altogether when the mass becomes of the order of $m^2{\sim}10$.  At this stage the well in the specific heat becomes on the other hand deeper and more pronounced and its minimum $\tilde{\alpha}_{mi}$ is moving  slowly to smaller values of the coupling constant  $\tilde{\alpha}$.  By inspection of the data  we can see that $\tilde{\alpha}_{mi}$ and  ${\alpha}_{s}$ starts to agree for larger masses and thus $\tilde{\alpha}_{mi}$ captures exactly the ${\bf S}^2_N$-to-matrix phase transition in this regime. The physical meaning of the the critical point  $\tilde{\alpha}_{ma}$ becomes also different for large masses where it becomes significantly different from ${\alpha}_{s}$. Since there is no peak the definition of $\tilde{\alpha}_{ma}$ becomes different\footnote{ It is not difficult to see that this new definition is  consistent with the previous definition of $\tilde{\alpha}_{ma}$. }. $\tilde{\alpha}_{ma}$   is now the value  of $\tilde{\alpha}$ at which the specific heat jumps and becomes equal to $N^2$. This is where the one-plaquette phase transition between weak and strong regimes of gauge theory on the fuzzy sphere occurs.  In figure $3$ we plot the specific heat versus $\hat{\alpha}$ for $N=10,12,16$ and $m^2=0.25,3,100$. In particular remark how the shape of the specific heat changes with $m$.


As it turns out we can predict the  ${\bf S}^2_N$-to-matrix phase transition from the one-loop theory of the model (\ref{model}).  To this end we consider the following background matrices $D_a=\alpha \phi L_a$ where $\phi$ is the radius of the sphere and $L_a$ are the generators of $SU(2)$ in the irreducible representation $\frac{N-1}{2}$. Then we compute the one-loop effective potential in the background field method \cite{ref} or by using an RG method \cite{ref1}. One finds the result
 
\begin{eqnarray}
V_{1-loop}&=&\frac{N^2\tilde{\alpha}^4}{2}\bigg[\frac{1+m^2}{4}{\phi}^4-\frac{1}{3}{\phi}^3-\frac{m^2}{2}{\phi}^2\bigg]+N^2\log\phi +O(N).\label{emp}
\end{eqnarray}
It is not difficult to check that the corresponding equation of motion of the potential (\ref{emp}) admits two real solutions where we can identify the one with the least energy  with the actual radius of the sphere. This however is only true up to a certain value $\tilde{\alpha}_{*}$ of the coupling constant $\tilde{\alpha}$ where the quartic equation ceases to have any real solution and as a consequence the fuzzy sphere solution $D_a=\alpha \phi L_a$ ceases to exist. In other words the potential below the value $\tilde{\alpha}_{*}$ of the coupling constant becomes unbounded and the fuzzy sphere  collapses. The critical values can be easily computed and one finds in the limit $m{\longrightarrow}0$ the values ${\phi}_{*}=0.75$ and $\tilde{\alpha}_{*}=2.09$. Extrapolating to large masses  we obtain the scaling behaviour 
\begin{eqnarray}
{\phi}_{*}=\frac{1}{\sqrt{2}}
\end{eqnarray}
and 
\begin{eqnarray}
\tilde{\alpha}_{*}=\big[\frac{8}{m^2+\sqrt{2}-1}\big]^{\frac{1}{4}}. \label{pred}
\end{eqnarray}
In other words the phase transition happens each time at a smaller value of the coupling constant $\tilde{\alpha}$ and thus the fuzzy sphere is more stable. This one-loop result is compared to the non-perturbative results ${\alpha}_s$ and ${\alpha}_{mi}$ coming from  the Monte Carlo simulation of the model (\ref{model}) in figure $4$. As one can immediately see there is an excellent agreement between the three values in the regime of large masses as discussed above. $\tilde{\alpha}_{*}$ and ${\alpha}_s$ agree as well for small masses. See also the phase diagram $(1)$.

For large values of $m$ the scaling of the coupling constant $\tilde{\alpha}$ as well as of the mass parameter $m$ is found to differ considerably from the $m=0$ case. It is now given by $
\bar{\alpha}=\tilde{\alpha}\sqrt{N}$, $\bar{m}=\frac{m}{N}$
The above theoretical fit (\ref{pred}) will read in terms of $\bar{\alpha}$ and $\bar{m}$ as follows
\begin{eqnarray}
\bar{\alpha}_{*}=\big[\frac{8}{\bar{m}^2}\big]^{\frac{1}{4}}.\label{5}
\end{eqnarray}
This is the fit used for the lower critical line in the phase diagram $(1)$. 
The critical value $\bar{\alpha}_s={\alpha}_s\sqrt{N}$ falls nicely on the top of this fit for all values of the mass.

The fit of the critical value $\tilde{\alpha}_{ma}$ for $m$ small is given by  equation (\ref{pred}). Thus we expect agreement between $\tilde{\alpha}_{*}$ and ${\alpha}_s$ from one hand  and $\tilde{\alpha}_{ma}$ from the other hand in the range of small masses.  For $m$ large we find that we can fit the data  $\tilde{\alpha}_{ma}$ to ( recall that ${\alpha}_p=\tilde{\alpha}_{ma}\sqrt{N}$ )
\begin{eqnarray}
{\alpha}_{p}=3.35{\pm}0.25+\big[\frac{0.04}{\bar{m}^2}\big]^{\frac{1}{2}}.\label{fitintr}
\end{eqnarray}
In other words in the limit $m{\longrightarrow}{\infty}$ we can fit the data to the line ${\alpha}_p=3.35$. This is what we call the one-plaquette critical line. See figure $5$. This is the fit used for the upper critical line in the phase diagram $(1)$. 



\begin{figure}
\begin{center}
\includegraphics[width=7cm,angle=-90]{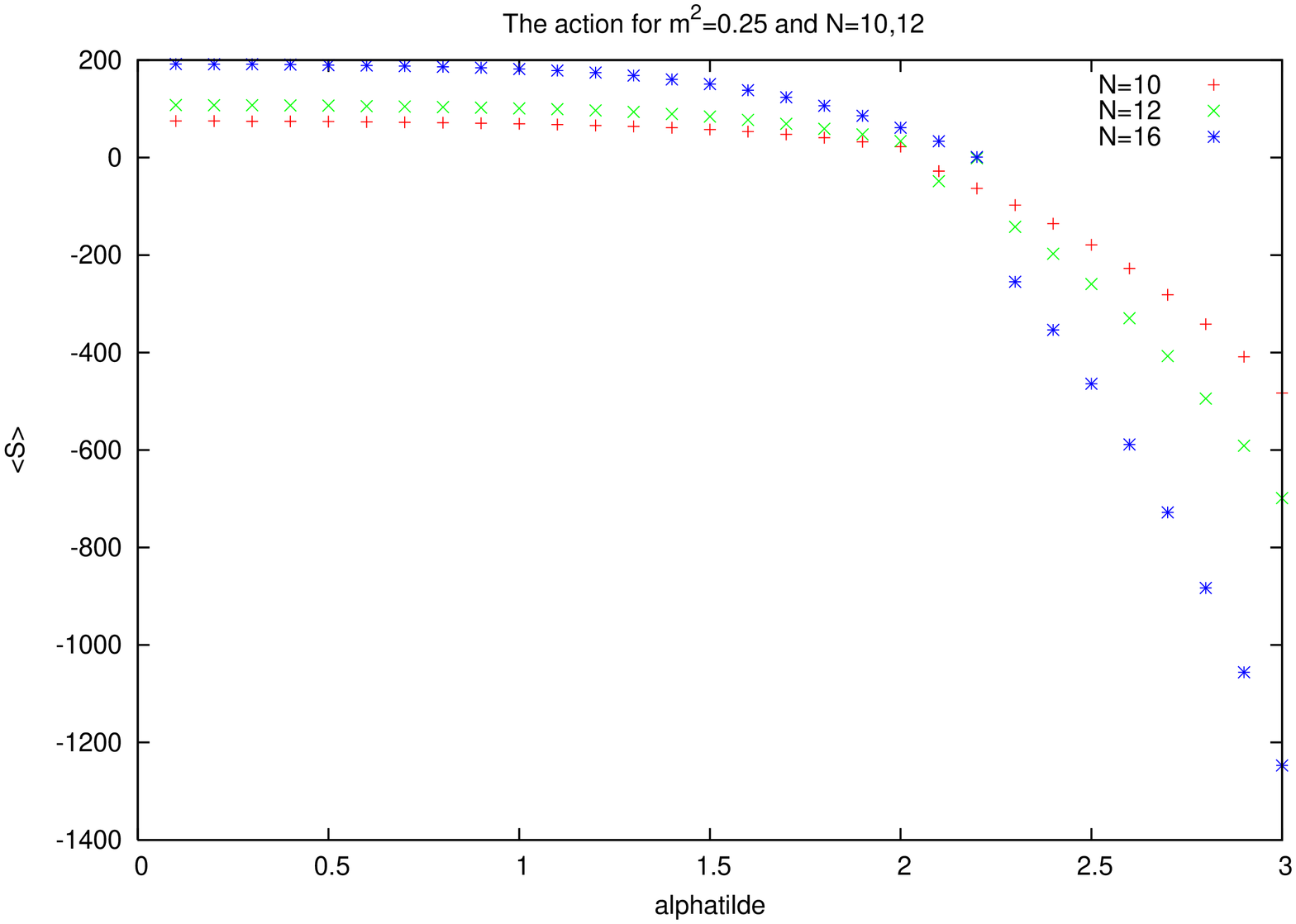} 
\includegraphics[width=7cm,angle=-90]{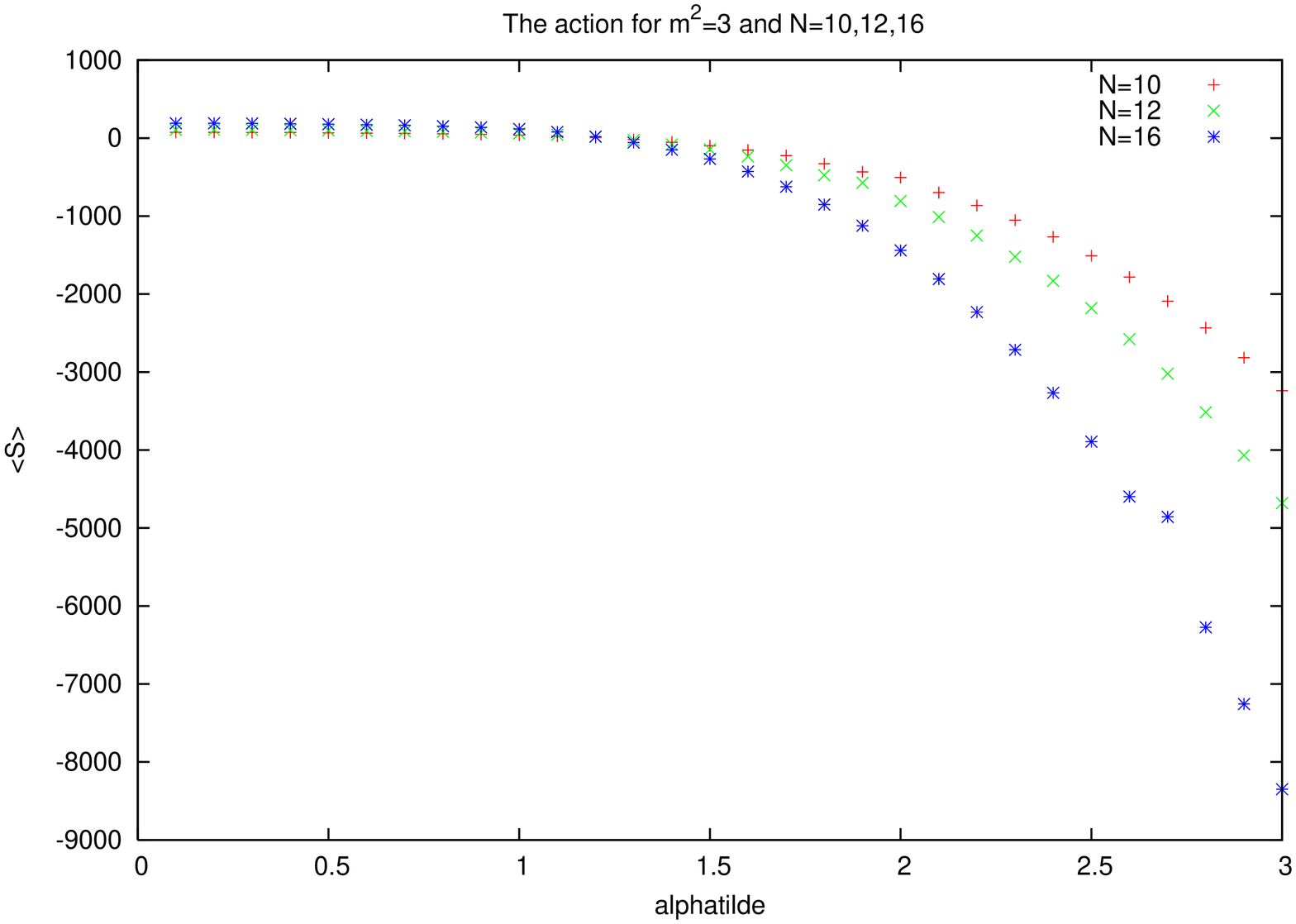}
\includegraphics[width=7cm,angle=-90]{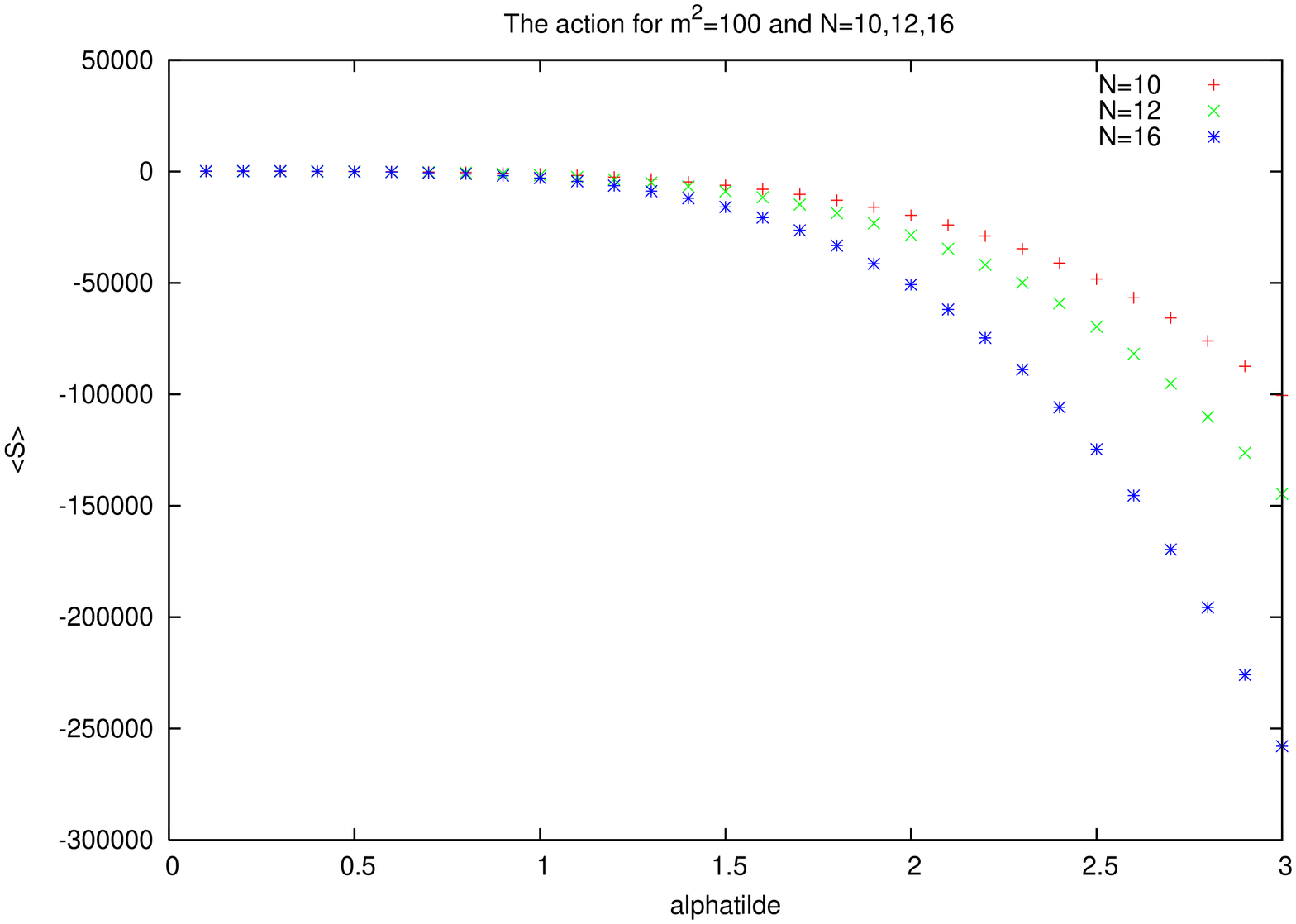} 
\caption{{ The action for  $m^2=0.25,3,100$  and $N=10,12,16$.}}
\end{center}
\end{figure}

\begin{figure}
\begin{center}
\includegraphics[width=7cm,angle=-90]{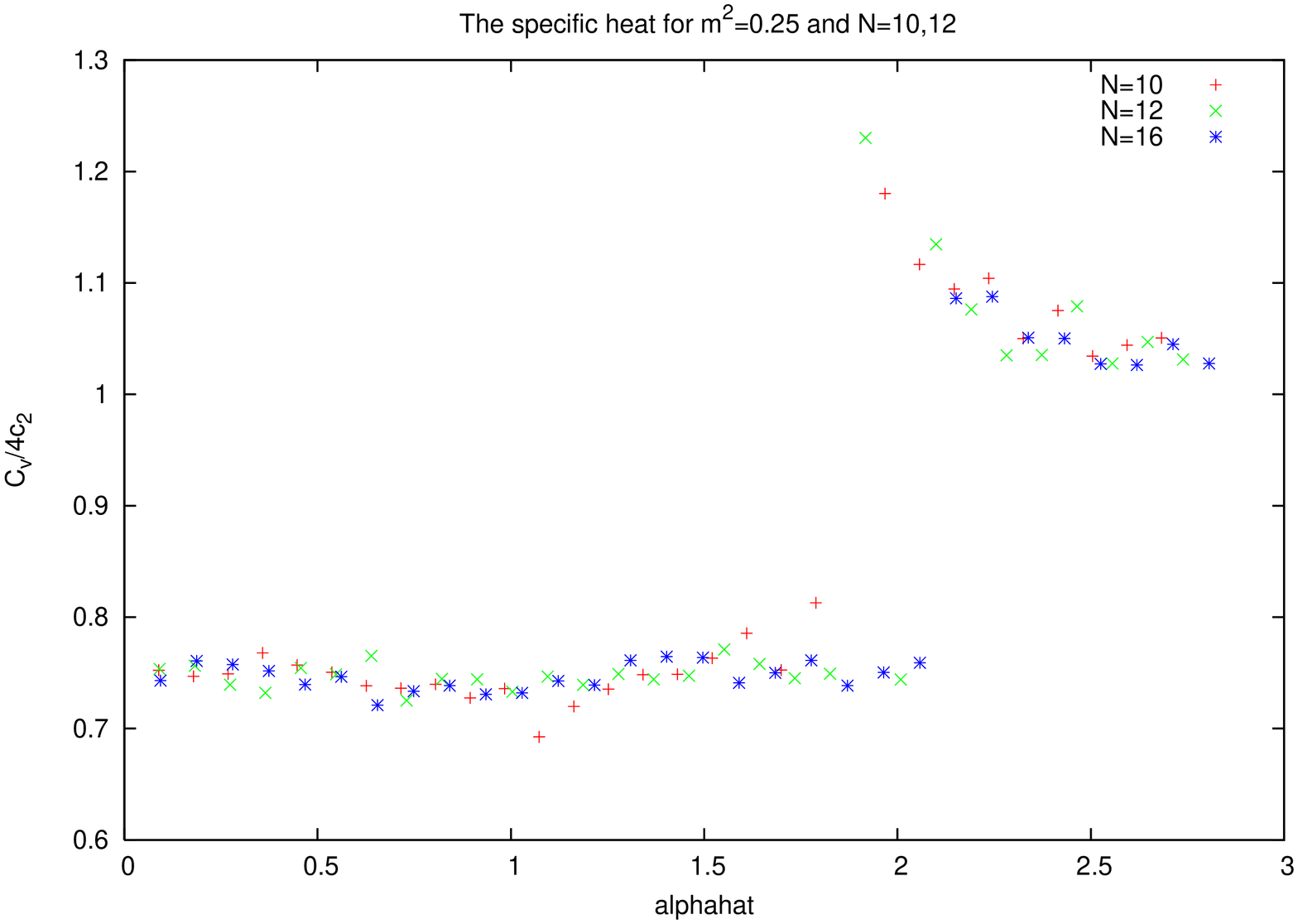}
\includegraphics[width=7cm,angle=-90]{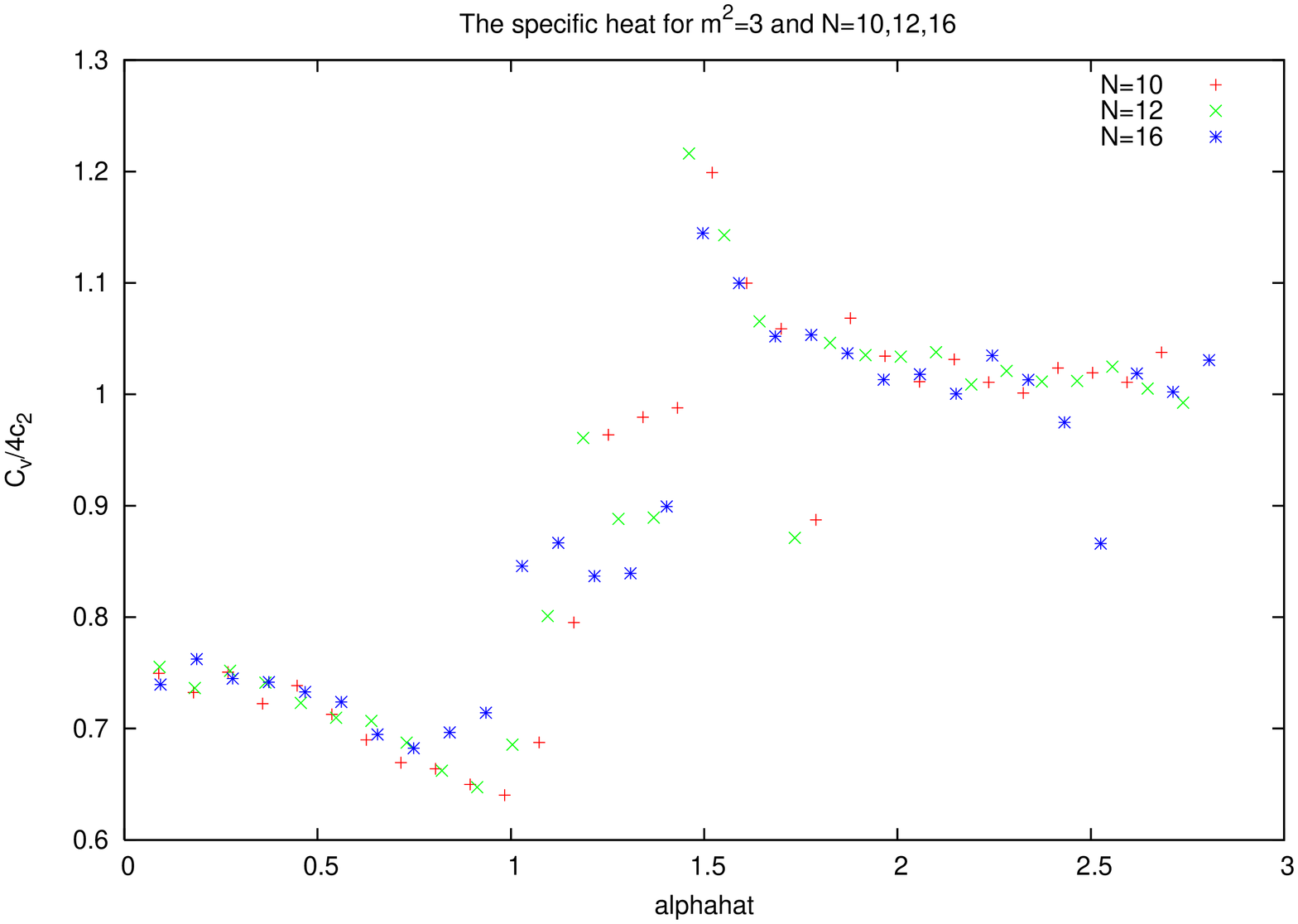}
\includegraphics[width=7cm,angle=-90]{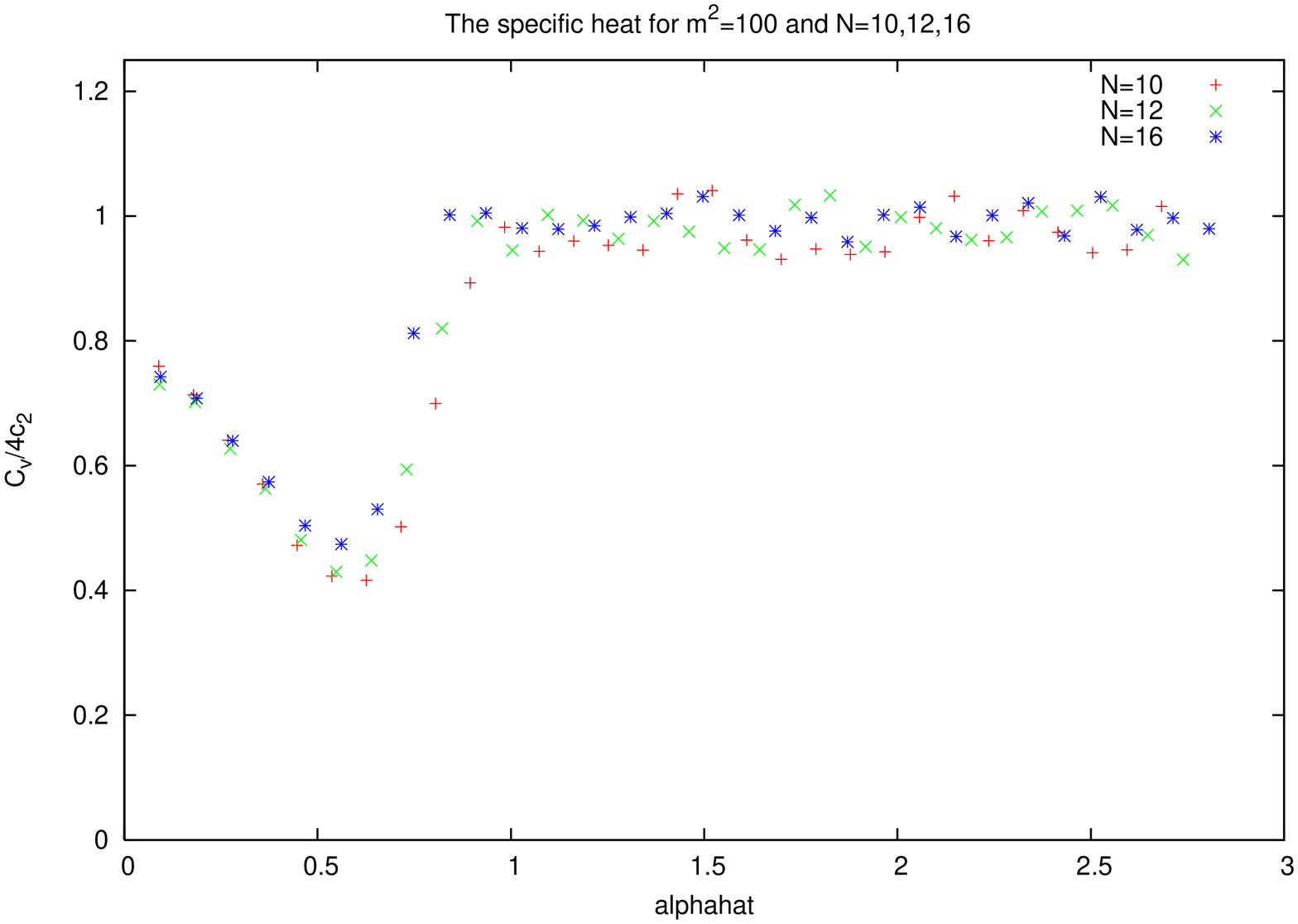}
\caption{{ The specific heat for $m^2=0.25,3,100$ and $N=10,12,16$.}}
\end{center}
\end{figure}

\begin{figure}
\begin{center}
\includegraphics[width=7cm,angle=-90]{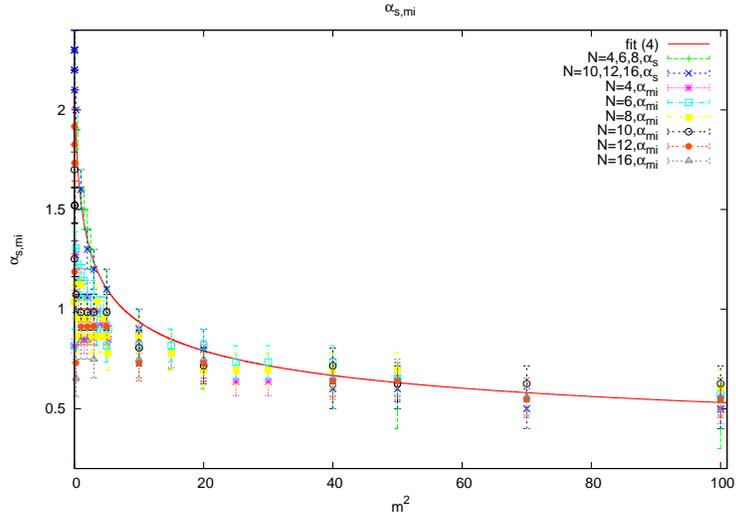}
\caption{{The ${\bf S}^2_L-$to-matrix critical line.}}
\end{center}
\end{figure}

\begin{figure}
\begin{center}
\includegraphics[width=7cm,angle=-90]{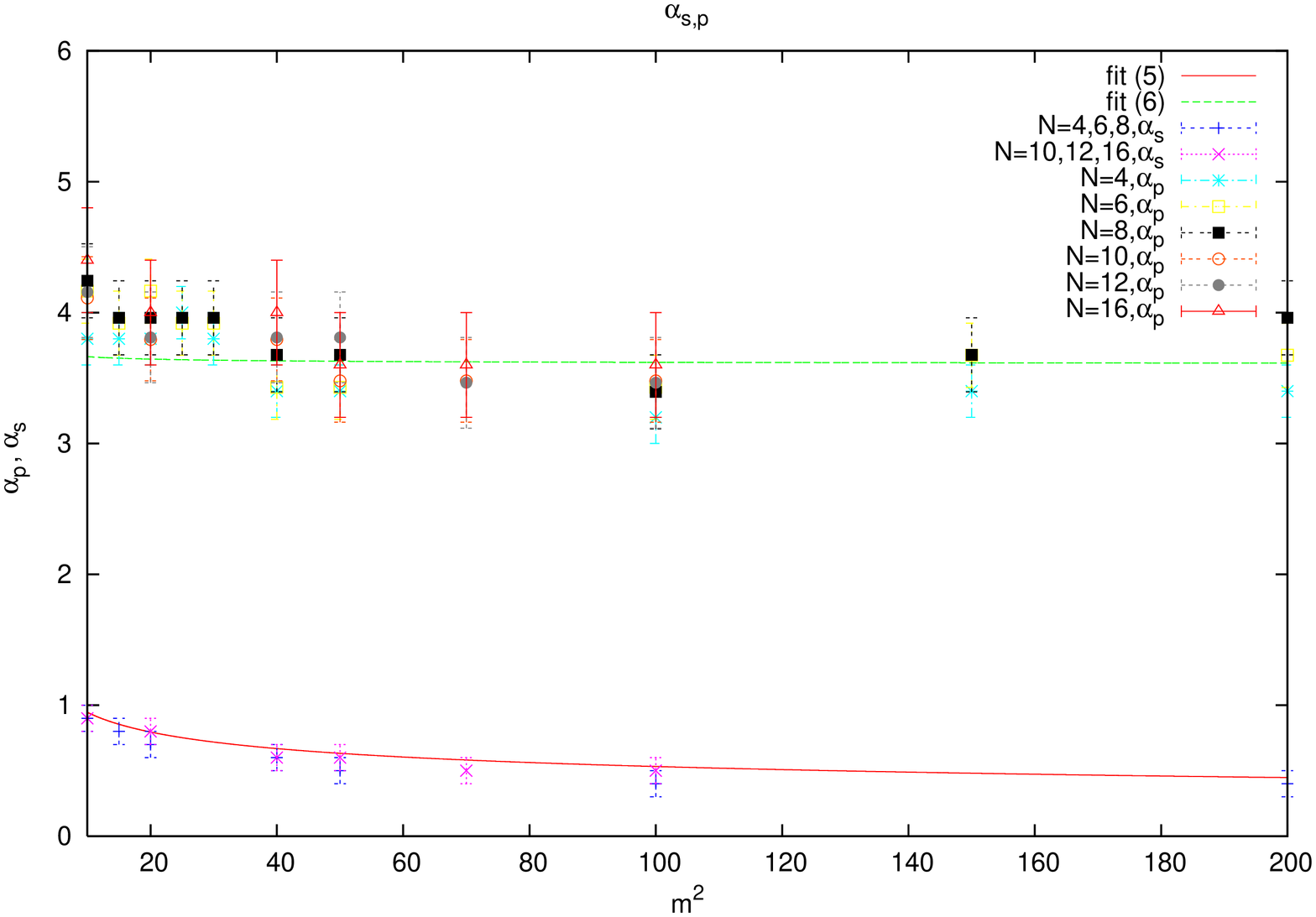}
\caption{{The one-plaquette critical line.}}
\end{center}
\end{figure}

\section{The one-loop calculation}
In this section we will follow \cite{ref}.

We are interested in the most general gauge theory up to quartic power in the gauge field on the fuzzy sphere ${\bf S}^2_{L+1}$. This is obtained as follows. Let $X_a$ , $a=1,2,3$, be three $N{\times}N$ hermitian matrices and let us consider the action
\begin{eqnarray}
S=N\bigg[-\frac{1}{4}Tr[X_a,X_b]^2+\frac{2i{\alpha}}{3}{\epsilon}_{abc}TrX_aX_bX_c\bigg]+\beta TrX_a^2+M Tr(X_a^2)^2.\label{action}
\end{eqnarray}
This action is invariant under the unitary transformations $X_a{\longrightarrow}UX_aU^{+}$. This model is also invariant under $SU(2)$ rotational symmetries $X_a{\longrightarrow}gX_ag^{+}=R_{ab}(g)X_b$ where the group element $g$ is given explicitly by $g=exp(i{\omega}_aL_a)$ for some constant vector $\vec{\omega}$ and $R(g)$ is the spin one irreducible representation of $g$. $L_a$ are the generators of $SU(2)$ in the irreducible representation $\frac{L}{2}$. They satisfy $[L_a,L_b]=i{\epsilon}_{abc}L_c$, $L_a^2=c_2=\frac{L}{2}(\frac{L}{2}+1)$ and  they are of size $(L+1){\times}(L+1)$.  This action is bounded from below for all positive values of $M$ and the trace is normalized such that $Tr1=N=L+1$. 

The $\alpha$, $\beta$ and $M$ are the parameters of the model. We are interested in the particular case where $\beta=-2Mc_2{\alpha}^2$. In this case the potential becomes
\begin{eqnarray}
V=\beta TrX_a^2+M Tr(X_a^2)^2=MTr(X_a^2-c_2{\alpha}^2)^2-MN{\alpha}^4c_2^2.\label{pote}
\end{eqnarray}
In most of this article we will discuss $U(1)$ gauge theory on the fuzzy sphere. We start with the values  $M=\beta=0$. This corresponds to the Alekseev, Recknagel, Schomerus action obtained in effective string theory describing the dynamics of open strings moving in a curved background with ${\bf S}^3$ metric in the presence of a Neveu-Schwarz B-field. We notice that with $M=\beta=0$ the trace part of $X_a$ simply decouples. As a consequence we can take $X_a$ to be traceless without any loss of generality. The classical absolute minimum of the model is given by
\begin{eqnarray}
X_a=\alpha L_a \label{fuzz}
\end{eqnarray}
where $L_a$ are the generators of $SU(2)$ in the irreducible representation $\frac{L}{2}{\equiv}\frac{N-1}{2}$.The quantum minimum is found by considering the configuration $X_a=\alpha {\phi}L_a$ 
where the order parameter $\alpha \phi$ plays the role of the radius of the sphere with a classical value equal $\alpha$.The complete one-loop effective potential in this configuration is given in the large $N$ limit by the  formula ( with $\tilde{\alpha}=\sqrt{N}{\alpha}$ ) 
\begin{eqnarray}
&&V_{\rm eff}(\phi)=2c_2\tilde{\alpha}^4 \bigg[\frac{1}{4}{\phi}^4-\frac{1}{3}{\phi}^3\bigg]+4c_2\log{\phi}+{\rm subleading~terms}.\label{formula}
\end{eqnarray}
It is not difficult to check that the equations of motion admits two real solutions where we can identify the one with the least energy  with the actual radius of the sphere. However this is only true up to a certain value $\tilde{\alpha}_{*}$ of the coupling constant $\tilde{\alpha}$ where the quartic equation ceases to have any real solution and as a consequence the fuzzy sphere solution (\ref{fuzz}) ceases to exist. In other words the potential $V_{\rm eff}$ below the value $\tilde{\alpha}_{*}$ of the coupling constant becomes unbounded and the fuzzy sphere  collapses. The critical value can be easily computed and one finds
\begin{eqnarray}
&&{\phi}_{*}=\frac{3}{4}~,~\tilde{\alpha}_{*}=2.08677944.
\end{eqnarray}
Now we add the potential term (\ref{pote}) with mass parameter $2M={Nm^2}/{c_2}$. In this case the matrices $X_a$ can not be taken traceless. The effective potential becomes
\begin{eqnarray}
V_{\rm eff}=2c_2\tilde{\alpha}^4 \bigg[\frac{1}{4}{\phi}^4-\frac{1}{3}{\phi}^3+\frac{1}{4}m^2({\phi}^2-1)^2\bigg]+4c_2\log{\phi}+\frac{1}{2}Tr_3TRlog{\Delta}.
\end{eqnarray}
$TR$ is the trace over  $4$ indices corresponding  to the left and right actions of operators on matrices of size $L+1$ while $Tr_3$ is the trace associated with the action of $3-$dimensional rotations. The Laplacian ${\Delta}$ in the gauge ${\xi}^{-1}=1+\frac{m^2}{c_2}$ is given by 
\begin{eqnarray}
{\Delta}&=&{\cal L}^2+(\frac{1}{\phi}-1)(J^2-{\cal L}^2-2)+2m^2(1-\frac{1}{{\phi}^2}).
\end{eqnarray}
The eigenvalues of ${\cal L}^2$ ( which is the Laplacian on the sphere ) and $\vec{J}^2$ ( which is the total angular momentum on the sphere ) are given respectively 
by $l(l+1)$ and $j(j+1)$ where $l=1,...,L$ and $j=l+1,l,l-1$. The corresponding eigentensors are the vector spherical harmonics operators. Let us also notice  that from the requirement that the spectrum of ${\Delta}$ must be positive we can derive a lower and upper bounds on the possible values which the field $\phi$ can take. For example for $m^2=0$ we can find that $2/3<\phi<3$.

We can show ( at least ) for small values of the mass $m$ that the logarithm of ${\Delta}$ is subleading in the large $N$ limit compared to the other terms and thus the potential reduces to the simpler form
 \begin{eqnarray}
V_{\rm eff}=2c_2\tilde{\alpha}^4 \bigg[\frac{1}{4}{\phi}^4-\frac{1}{3}{\phi}^3+\frac{1}{4}m^2({\phi}^2-1)^2\bigg]+4c_2\log{\phi}.
\end{eqnarray}
Solving for the critical value using the same method outlined previously yields the results
\begin{eqnarray}
{\phi}_{*}=\frac{3}{8(1+m^2)}\bigg[1+\sqrt{1+\frac{32m^2(1+m^2)}{9}}\bigg].\label{pre}
\end{eqnarray}
\begin{eqnarray}
\frac{1}{\tilde{\alpha}^4_{*}}=-\frac{1}{2}(1+m^2){\phi}_{*}^4+\frac{1}{2}{\phi}_{*}^3+\frac{m^2}{2}{\phi}_{*}^2.\label{pre1}
\end{eqnarray}
Extrapolating to large masses  ($m{\longrightarrow}{\infty}$) we obtain the scaling behaviour 
\begin{eqnarray}
\tilde{\alpha}_{*}=\big[\frac{8}{m^2+\sqrt{2}-1}\big]^{\frac{1}{4}}. \label{pre2}
\end{eqnarray}
In other words the phase transition happens each time at a smaller value of the coupling constant $\tilde{\alpha}$ and thus the fuzzy sphere is more stable.

It is therefore sensible to expand the action (\ref{action}) around the fuzzy sphere solution (\ref{fuzz}) by introducing a $U(1)$ gauge field $A_a$ on the fuzzy sphere ${\bf S}^2_N$ as follows $X_a={\alpha}(L_a+A_a)$. The action becomes 
\begin{eqnarray}
S_N&=&\frac{1}{4g^2N}Tr\left[F_{ab}^{(0)}+i[A_a,A_b]\right]^2-\frac{1}{2g^2N}{\epsilon}_{abc}Tr\left[\frac{1}{2}F_{ab}^{(0)}A_c+\frac{i}{3}[A_a,A_b]A_c\right]+\frac{2m^2}{g^2N}Tr \Phi^2\nonumber\\
&-&\frac{1}{6}\tilde{\alpha}^4c_2-\frac{1}{2}\tilde{\alpha}^4c_2m^2.
\end{eqnarray}
$\Phi$ is the normal covariant scalar component of the gauge field on the fuzzy sphere defined by $\sqrt{4c_2} \Phi=L_aA_a+A_aL_a+A_a^2$ . $F_{ab}=F_{ab}^{(0)}+i[A_a,A_b]$ is the $U(1)$ covariant curvature where $F_{ab}^{(0)}=i[L_a,A_b]-i[L_b,A_a]+{\epsilon}_{abc}A_c$ and $g$ is the gauge coupling constant defined by $\frac{1}{g^2}=\tilde{\alpha}^4$. In the continuum limit $L{\longrightarrow}{\infty}$ all commutators vanish and we get a $U(1)$ gauge field coupled to a scalar mode $\Phi=\vec{n}.\vec{A}$ with curvature $F_{ab}^{(0)}=i{\cal L}_aA_b-i{\cal L}_bA_a+{\epsilon}_{abc}A_c$ where ${\cal L}_a=-i{\epsilon}_{abc}n_b\frac{\partial}{{\partial}n_c}$. We find ( by also neglecting the constant term )
\begin{eqnarray}
S_{\infty}&=&\frac{1}{4g^2} \int_{S^2} \frac{d{\Omega}}{4{\pi}}(F_{ab}^{(0)})^2-\frac{1}{4g^2}{\epsilon}_{abc}\int_{S^2}\frac{d{\Omega}}{4\pi}F_{ab}^{(0)}A_c+\frac{2m^2}{g^2}\int_{S^2}\frac{d{\Omega}}{4\pi} \Phi^2.
\end{eqnarray}
The quantization of the fuzzy action $S_N$  yields a non-trivial effective action ${\Gamma}_{\infty}$ in the continuum limit which for generic values of the mass parameter $m$ is different from $S_{\infty}$.   For example we have established by explicit calculation of the quadratic effective action the existence of a gauge-invariant UV-IR mixing problem in $U(1)$ gauge theory on the fuzzy sphere for the value $m=0$. We find
\begin{eqnarray}
{\Gamma}_{\infty}^{(2)}&=& \frac{1}{4g^2}\int_{S^2}
\frac{d{\Omega}}{4{\pi}}F_{ab}^{(0)}(1+2g^2{\Delta}_3)F_{ab}^{(0)}-\frac{1}{4g^2}{\epsilon}_{abc}\int_{S^2}
\frac{d{\Omega}}{4{\pi}}F_{ab}^{(0)}(1+2g^2{\Delta}_3)A_c+4\sqrt{c_2}\int_{S^2}\frac{d{\Omega}}{4{\pi}}\Phi \nonumber\\
&+&{\rm other~non~local~ quadratic ~terms}.\label{main1}
\end{eqnarray}
The operator ${\Delta}_3$ is a  function of the Laplacian ${\cal L}^2$ with eigenvalues ${\Delta}_3(k)$ given by $k(k+1){\Delta}_3(k)=\sum_{p=2}^{k}\frac{1}{p}$. Clearly ${\Gamma}_{\infty}^{(2)}{\neq}S_{\infty}$ which is the signature of the UV-IR mixing in this model.It is expected that the same result will also hold for generic values of the mass parameter $m$. 

The calculation can also be done quite easily in the limit $m{\longrightarrow}{\infty}$ and one finds that there is no UV-IR mixing in the model in this case. The UV-IR mixing is thus confined to the scalar sector of the model since the limit $m{\longrightarrow}{\infty}$ projects out the scalar fluctuation $\Phi$. It is hence natural to think that the extra matrix phase observed in the phase structure of the theory is related to this mixing; in other words it is the non-perturbative manifestation of the perturbative UV-IR mixing property since as we have shown this phase seems also to disappear in the limit of large masses.

\section{Monte Carlo simulation}
From the above one-loop argument it is expected to observe at least one phase transition on the line $m=\beta=0$. This is a continuous first order phase transition from a fuzzy sphere phase with $\alpha > {\alpha}_{*}$ to a pure matrix phase with $\alpha <\alpha_{*}$. It is also expected that this critical value ${\alpha}_{*}$ decreases with the value of $m$ ( keeping $\beta$ fixed equal $-2Mc_2{\alpha}^2$ with $2M={Nm^2}/{c_2}$ ) and it becomes zero when we let $m{\longrightarrow}{\infty}$. 

\subsection{Zero mass}
We start with $M=\beta=0$. To detect the different phases of the model we propose to measure the following observables. First we measure the average value of the action, viz $<S>$. Second the specific heat will allow us to demarcate the boundary between the different phases. It is defined by $C_v=<S^2>-<S>^2$.

In order to determine the critical point (if any) we run several simulations with different values of $N$, say $N=4,6,8,10,12,16$. We always start from a random (hot) initial configuration and run the metropolis algorithm for ${\rm Ttherm}+{\rm Tcorr}{\times}{\rm Tmont}$ Monte Carlo steps.  ${\rm Ttherm}$ is thermalization time while ${\rm Tmont}$ is the actual number of Monte Carlo steps. Two consecutive Monte Carlo times are separated by ${\rm Tcorr}$ sweeps to reduce auto-correlation time. In every step ( sweep ) we go through each entry of the three matrices $X_1$, $X_2$ and $X_3$ and update it according to the Boltzman weight. This by definition is one unit of time ( Monte Carlo time ) in the generated dynamics. For every $N$ and ${\alpha}$ we tune appropriately ${\rm Ttherm}$, ${\rm Tmont}$, ${\rm Tcorr}$ as well as the interval $I$ from which we choose the variation of the entries of the matrices $X_a$ so that to reduce auto-correlation times and statistical errors. 

The continuum limit of a given observable will be obtained by collapsing the corresponding data, in other words finding the scaling of this operator in the large $N$ limit which yields an $N-$independent quantity. For example the scaling of the coupling constant $\alpha$ with $N$ is clearly given by $\tilde{\alpha}=\sqrt{N}{\alpha}$ as anticipated from the one-loop calculation.

For the action the data is plotted in figure $6$. We remark that the $4$ curves with $N=4,6,8$ and $10$ all intersect around the point 
\begin{eqnarray}
{\alpha}_{s}=2.2{\pm}0.1.\label{31}
\end{eqnarray}
This is the critical point since it is independent of $N$ as it should be. The collapse of the data is given by  $<S>/4c_2$. Indeed a very good fit for the action $<S>$ is given by  the classical action in the configuration $X_a=\alpha L_a$, i.e
\begin{eqnarray}
<S>=-\frac{\tilde{\alpha}^4c_2}{6}
\end{eqnarray}
The data for the specific heat is shown on  figure $7$. We can immediately remark that $C_v$ peakes around the above critical point. More precisely the peak is at the values $\tilde{\alpha}=2.25{\pm}0.05,2.1{\pm}0.1$ and $2.1{\pm}0.1,2.2{\pm}0.1$ for $N=4,6$ and $8,10$ respectively.

From figure $7$ the correct scaling of the specific heat is given by $C_v/4c_2$. Let us also remark that the specific heat is equal $C_v=N^2-1$ in the fuzzy sphere phase and $C_v=0.75 (N^2-1)$ in the matrix phase.

\begin{figure}
\begin{center}
\includegraphics[width=7cm,angle=-90]{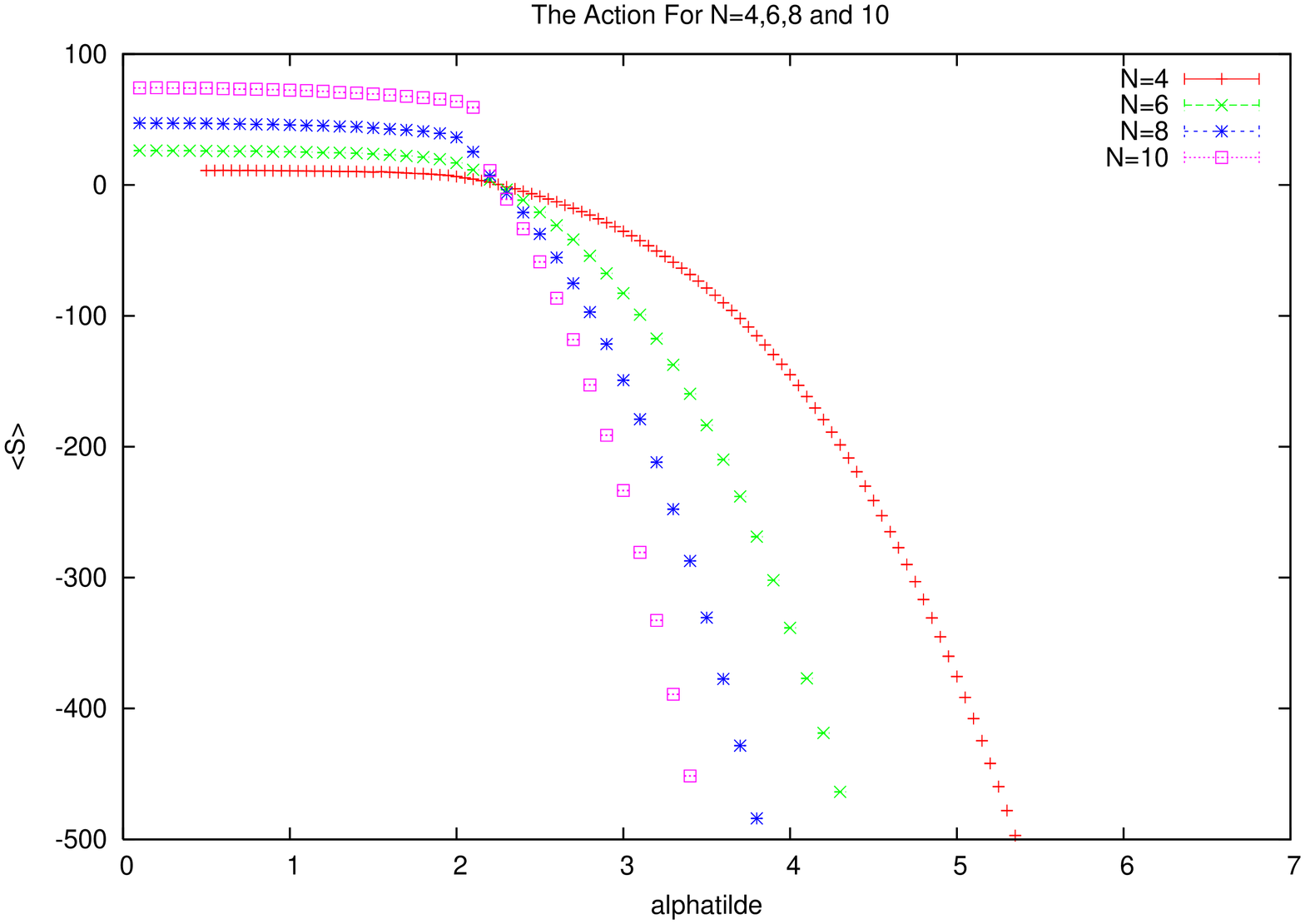} 
\includegraphics[width=7cm,angle=-90]{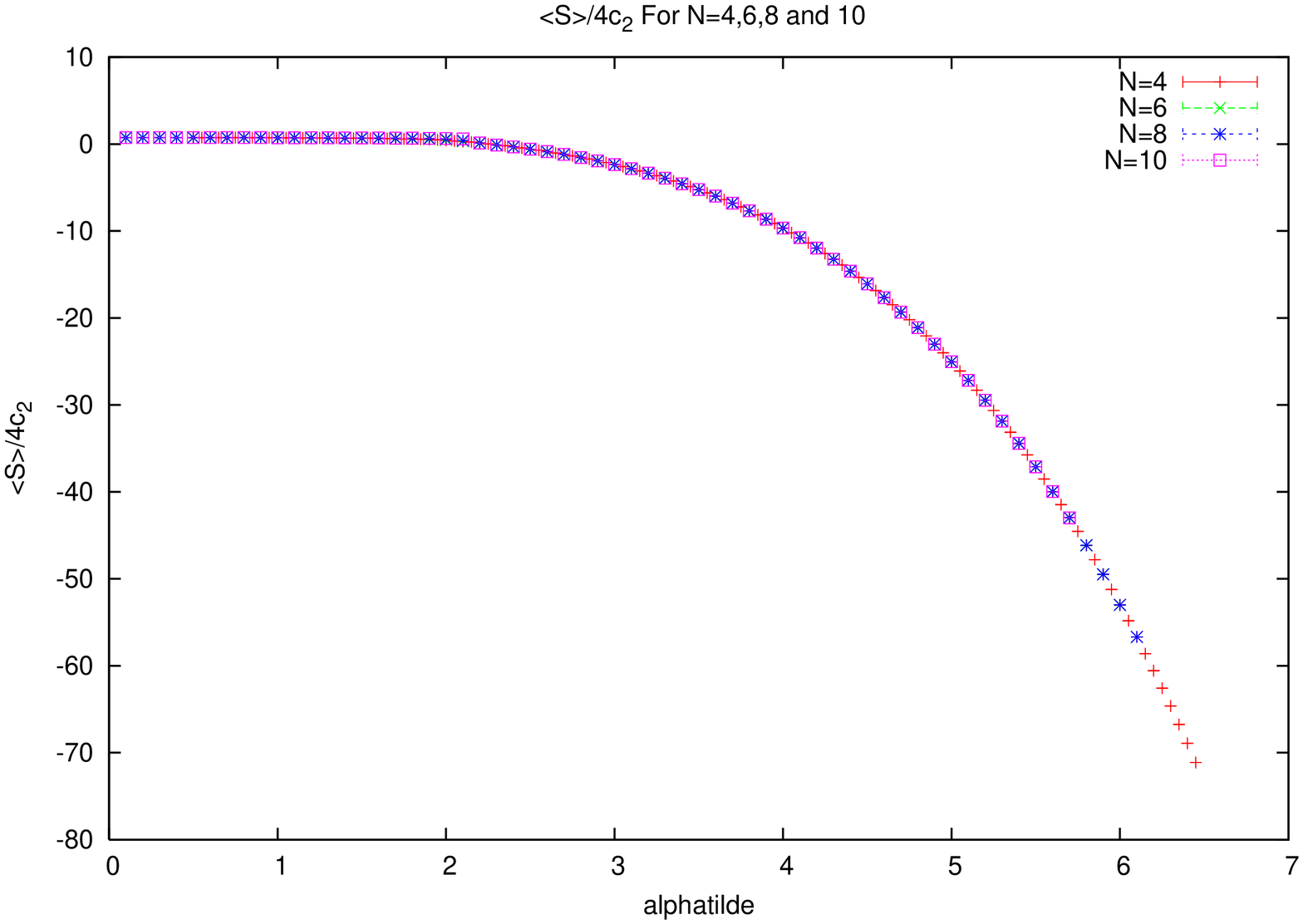}
\caption{{ The action for zero mass for $N=4,6,8,10$.}}
\end{center}
\end{figure}

\begin{figure}
\begin{center}
\includegraphics[width=7cm,angle=-90]{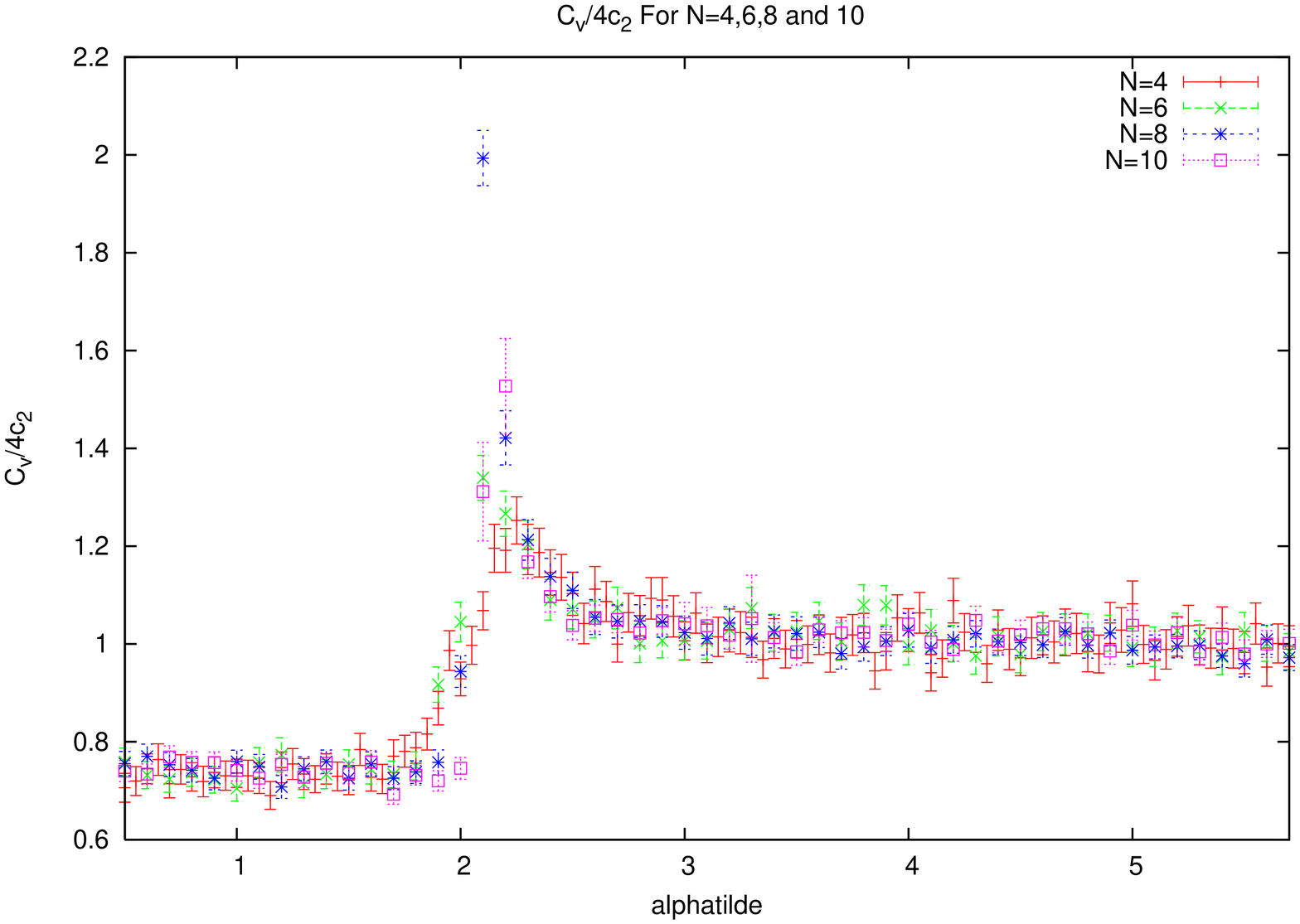} 
\caption{{ The specific heat for zero mass for $N=4,6,8,10$.}}
\end{center}
\end{figure}
We have therefore established the existence of a first order phase transition from the fuzzy sphere to a matrix phase in agreement with the one-loop calculation. The next step is to add to the Alekseev-Recknagel-Schomerus model the potential term (\ref{pote}).
\subsection{Non-zero mass : the ${\bf S}^2_N-$to-matrix phase transition}
 As we have shown $m^2$ plays precisely the role of the mass parameter of the normal scalar field in the fuzzy sphere phase. From the one-loop calculation as well as from the large $1/N$ expansion it is argued that the  fuzzy sphere becomes more dominant ( i.e it becomes more stable under quantum fluctuations ) as we increase the mass $m$ of the scalar mode. In the limit  $m{\longrightarrow}{\infty}$ we expect the matrix phase to disappear altogether. In this limit $m{\longrightarrow}{\infty}$ the normal scalar field decouples from the pure two dimensional gauge sector and as a consequence it is natural to conjecture  that the matrix phase ( and correspondingly the perturbative UV-IR mixing phenomena ) has its origin in the coupling of this extra normal mode to the rest of the dynamics. Another way of putting this conjecture is that the presence of the matrix phase ( which is absent in the continuum theory ) is nothing else but a non-perturbative manifestation of the perturbative UV-IR mixing. However although this is true to a large extent there are more non-trivial things happening in this limit as we will now report.

We again measure the action ${<S>}$ for non-zero values of the mass $m$ for $N=4,6$ and $8$. The results are shown on figure $8$. As before the action $<S>$ is scaled as $<S>/4c_2$ as a function of $\tilde{\alpha}$. It is also immediately clear that the critical point decreases as we increase $m$. In other words the fuzzy sphere becomes more dominant as promised by the one-loop calculation. For example for $m^2=0.5$ the  actions for $N=4$, $N=6$ and $N=8$ intersect at ${\alpha}_{s}=1.9{\pm}0.1$. The theory predicts a critical value given by $\tilde{\alpha}_{*}=1.72$ which is reasonably close.

We repeat the above calculation for various values of the mass $m$. The intersection point of the actions with different $N$ defines the critical value ${\alpha}_{s}$.  This value follows to a good accuracy the one-loop prediction given by equation (\ref{pre2}). As it turns out this phase transition is also captured by the minimum of the specific heat ( more on this below ). The phase diagram of the fuzzy sphere-to-matrix phase transition is shown on figure $9$. A very good fit of $<S>$ is given by the classical action in the fuzzy sphere configuration. For non-zero mass this is given by the expression 
\begin{eqnarray}
<S>=-\frac{\tilde{\alpha}^4c_2}{6}-\frac{m^2}{2}c_2\tilde{\alpha}^4.\label{fitS}
\end{eqnarray}

\begin{figure}
\begin{center}
\includegraphics[width=7cm,angle=-90]{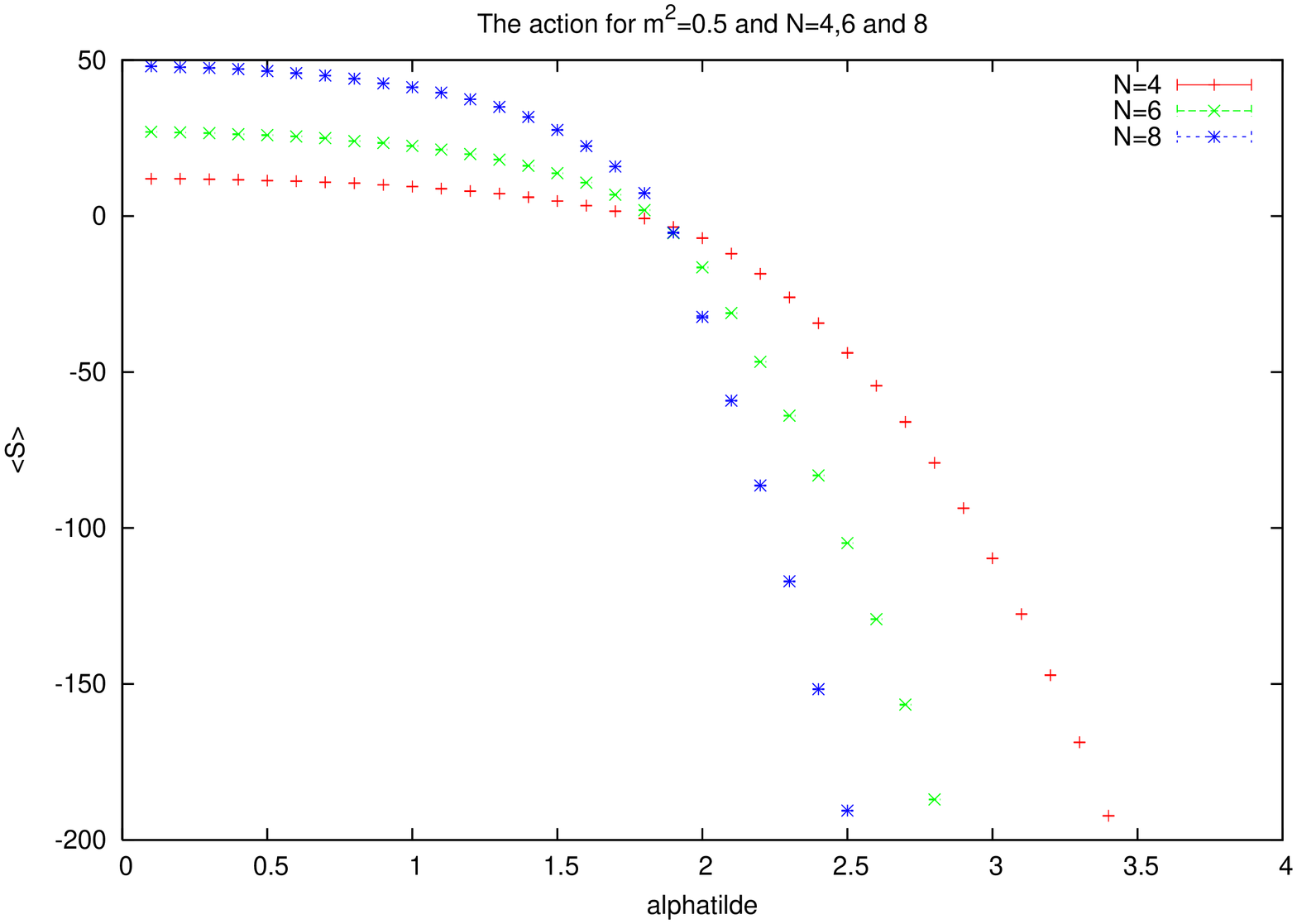}
\includegraphics[width=7cm,angle=-90]{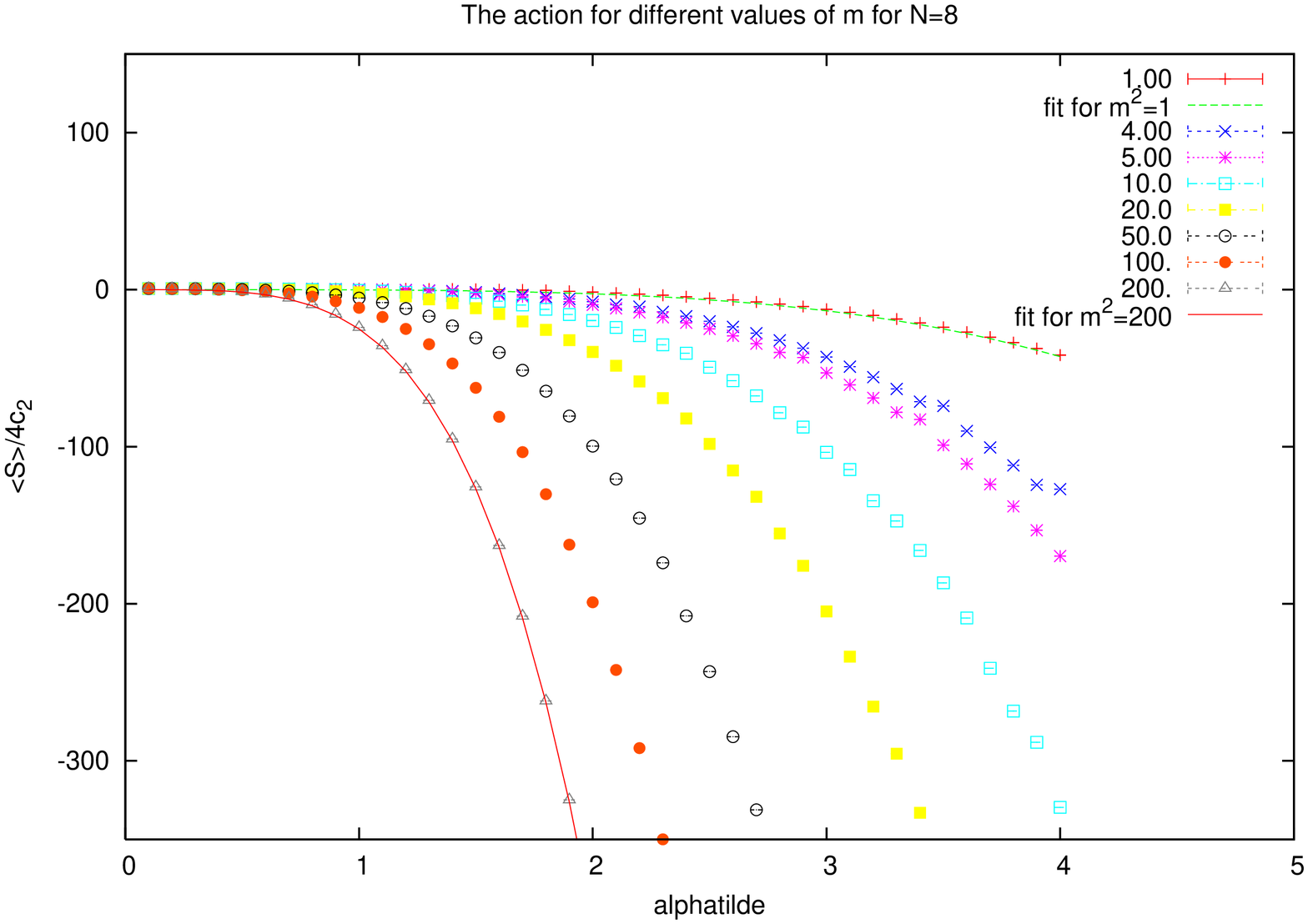}
\caption{{ The action for non-zero mass. The fit is given by equation (\ref{fitS}).}}
\end{center}
\end{figure}
\begin{figure}
\begin{center}
\includegraphics[width=13cm,angle=-90]{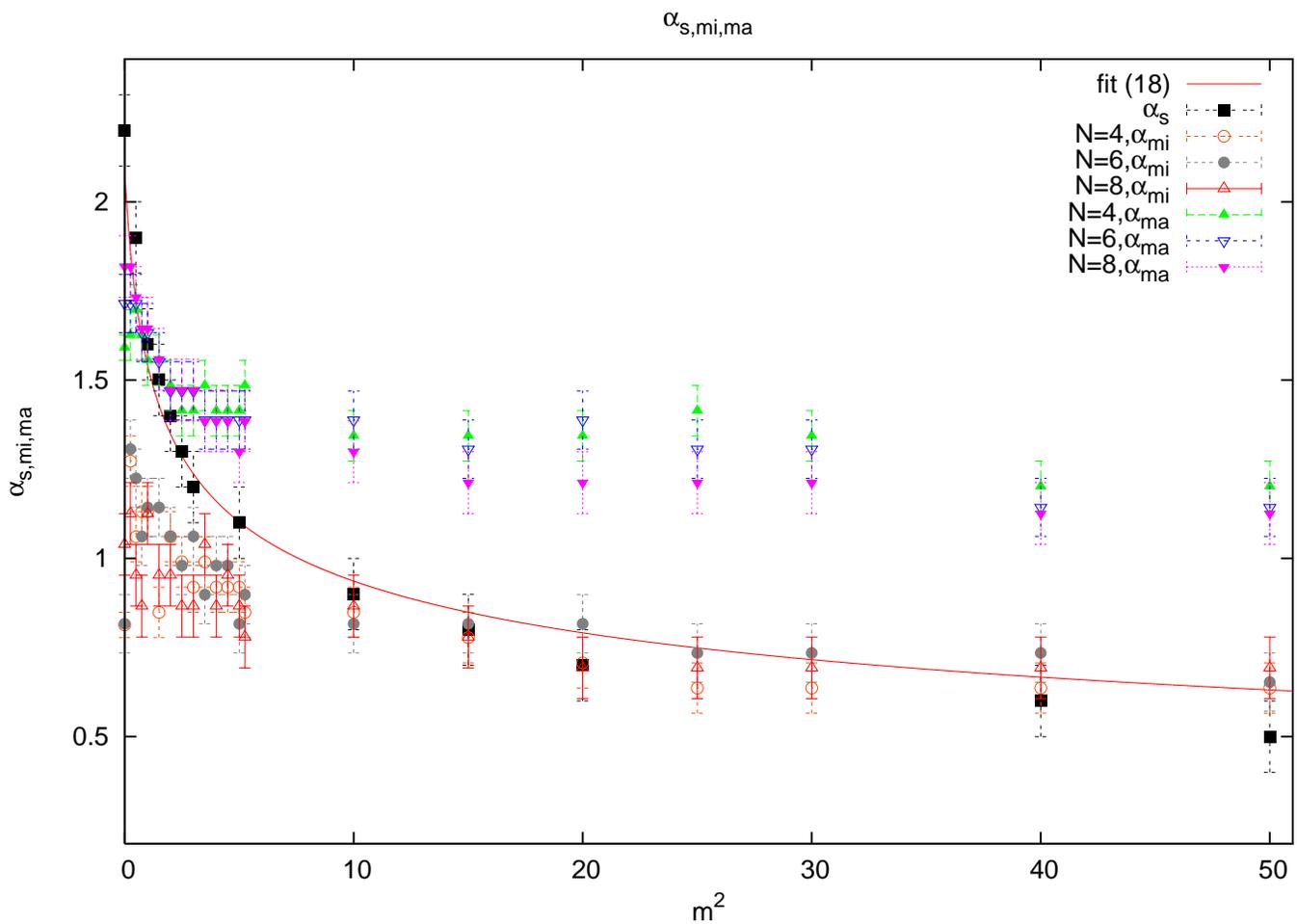}
\caption{{The phase diagram of the ${\bf S}^2_N-$to-matrix phase transition. The fuzzy sphere phase is above the solid line while the matrix phase is below it. The critical point ${\alpha}_{s}$ is the intersection point of the actions with $N=4,6$ and $8$. This  agrees with the one-loop prediction of the critical line ( the solid line ) given by equation (\ref{pre2}). For large masses ${\alpha}_{s}$ coincides with the minimum ${\alpha}_{mi}$ of $C_v$.  For small masses ${\alpha}_{s}$ coincides with the maximum ${\alpha}_{ma}$ of $C_v$. }}
\end{center}
\end{figure}

\subsection{Specific heat: the one-plaquette phase transition}

As soon as the mass $m$ takes a non-zero value the specific heat ${C}_{v}$ starts to behave in a very different way compared to its behaviour for zero mass. We observe a new phase transition for large enough masses which resembles very much the one-plaquette phase transition in ordinary $2-$dimensional gauge theory. This is measured by  the maximum ${\alpha}_{ma}$ of the specific heat. For small values of the mass parameter $m$ the maximum ${\alpha}_{ma}$ is defined by the position of the peak of $C_v$. For large values of $m$ the peak in $C_v$ disappears and   ${\alpha}_{ma}$ is given by the value of the coupling constant at which the specific heat discontinuously jumps to one. We will also measure the minimum ${\alpha}_{mi}$ of $C_v$ which will capture the ${\bf S}^2_N-$to-matrix phase transition.

For small values of $m$ the scaling of the coupling constant $\tilde{\alpha}$ is found to differ only slightly from the $m=0$ case. It is given by 
\begin{eqnarray}
\hat{\alpha}=\tilde{\alpha}\sqrt{1-\frac{2}{N}}. 
\end{eqnarray}
In the large $N$ limit $\hat{\alpha}/\tilde{\alpha}{\longrightarrow}1$ and thus this different scaling is only due to finite size effects and has no other physical significance. This is expected for small masses. ${\alpha}_{ma}$ and ${\alpha}_{mi}$ are actually the values of $\hat{\alpha}$ at the maximum and minimum of the specific heat.
The peak of the specific heat moves slowly  to smaller values of the coupling constant as we increase $m$. The agreement with the one-loop prediction given by equation (\ref{pre2}) as well as with ${\alpha}_s$ is fairly good and thus in this regime ${\alpha}_{ma}$ is still detecting the ${\bf S}^2_N-$to-matrix phase transition. Similarly to the case $m=0$ the specific heat  ${C}_{v}/4c_2$ as a function of $\hat{\alpha}$  is equal to $1$ in the fuzzy sphere phase. However there is a shallow valley starting to appear for values of $\hat{\alpha}$ inside the matrix phase. The values ${\alpha}_{mi}$ of the minimum of $C_{v}$ for these small masses are significantly different from ${\alpha}_s$ ( see the phase diagram on figure $9$ ). As an example the data for $m^2=0.25,4.75$ for $N=6,8$ is shown on figure $10$.

As $m$ keeps increasing deviation from the one-loop prediction becomes important. The data for $m^2=40,200$ for $N=6,8$ is shown on figure $10$. We observe that the peak flattens slowly and disappears altogether when the mass becomes of the order of $m^2{\sim}10$. 

Although the peak in $C_v$ disappears we know that the ${\bf S}^2_N$-to-matrix phase transition is still present as indicated by the non-vanishing of  ${\alpha}_{s}$ ( from the phase diagram on figure $9$ ) . The physical meaning of the the critical point  ${\alpha}_{ma}$ becomes different for large masses. This is where the one-plaquette phase transition between weak and strong regimes of gauge theory on the fuzzy sphere occurs. The valley in the specific heat becomes on the other hand deeper and more pronounced as $m$ increases and its minimum ${\alpha}_{mi}$ is moving  slowly to smaller values of the coupling constant  $\hat{\alpha}$.  By inspection of the data  ( phase diagram on figure $9$ ) we can see that ${\alpha}_{mi}$ and  ${\alpha}_{s}$ starts to agree for larger masses and thus ${\alpha}_{mi}$ captures exactly the ${\bf S}^2_N$-to-matrix phase transition.

The two regimes with $m$ small and $m$ large are thus physically distinct; in the first regime we have two phases : the fuzzy sphere phase $\tilde{\alpha}{\geq}{\alpha}_{s}$ and the matrix phase  $\tilde{\alpha}{\leq}{\alpha}_{s}$ whereas in the second regime we have three phases. Beside the matrix phase for $\hat{\alpha}{\leq}{\alpha}_{mi}$ we have two more phases where we have a stable fuzzy sphere as the underlying spacetime structure. These two phases correspond to $U(1)$ gauge theory on the fuzzy sphere ${\bf S}^2_N$ in the weak  $\hat{\alpha}{\geq}{\alpha}_{ma}$ and strong ${\alpha}_{mi}{\leq}\hat{\alpha}{\leq}{\alpha}_{ma}$ regimes respectively. There exists therefore a triple point where the three phases coexist.

For large values of $m$ the scaling of the coupling constant $\tilde{\alpha}$ as well as of the mass parameter $m$ is found to differ considerably from the $m=0$ case. It is now given by 
\begin{eqnarray}
\bar{\alpha}=\tilde{\alpha}\sqrt{N}~,~\bar{m}=\frac{m}{N}. 
\end{eqnarray}
The theoretical fit (\ref{pre2}) will read in terms of $\bar{\alpha}$ and $\bar{m}$ as follows
\begin{eqnarray}
\bar{\alpha}_{*}=\big[\frac{8}{\bar{m}^2}\big]^{\frac{1}{4}}.\label{pre3}
\end{eqnarray}
We define the one-plaquette transition point by the value ${\alpha}_{ma}$ of the coupling constant $\hat{\alpha}$ ( or equivalently $\tilde{\alpha}$ in the large $N$ limit ) at which the specific heat discontinuously jumps to one. In terms of $\bar{\alpha}$ this is given at the value
\begin{eqnarray}
{\alpha}_p={\alpha}_{ma}\sqrt{N}. 
\end{eqnarray}
The fit of the critical value ${\alpha}_{ma}$ for $m$ small is given by equation (\ref{pre2}) while for $m$ large we find that we can fit the data to
\begin{eqnarray}
{\alpha}_{p}=3.35{\pm}0.25+\big[\frac{0.04}{\bar{m}^2}\big]^{\frac{1}{2}}.\label{fit}
\end{eqnarray}
In other words in the limit $m{\longrightarrow}{\infty}$ we can fit the data to ${\alpha}_p=3.35{\pm}0.25$. In the next section we will give a theoretical derivation of the value $3.35$ from the one-plaquette approximation of gauge fields on the fuzzy sphere in the weak regime $\bar{\alpha}{\geq}{\alpha}_{p}$. 

These results are summarized in the phase diagram on figure $11$.

 Finally we point out that we can estimate the values $\bar{\alpha}_T$ and $\bar{m}_{T}^2$ of the coupling constant $\bar{\alpha}$ and the mass parameter $\bar{m}^2$ at the triple point by equating the fits (\ref{pre3}) and (\ref{fit}). We obtain the two solutions $1)$ $\bar{m}_{T}^2=0.009$ and  $\bar{\alpha}_T=5.46$ or equivalently  $\log\bar{m}_{T}^2=-4.71$ and  $\log\bar{\alpha}_T=1.7$ and $2)$ $\bar{m}_{T}^2=0.001$ and  $\bar{\alpha}_T=9.46$ or equivalently  $\log\bar{m}_{T}^2=-6.91$ and  $\log\bar{\alpha}_T=2.25$. The triple point must therefore exist between these two points, viz
\begin{eqnarray}
1.7{\leq}\log\bar{\alpha}_T{\leq}2.25 
\end{eqnarray}
and
\begin{eqnarray}
 -6.91{\leq}\log\bar{m}_{T}^2{\leq}-4.71.
\end{eqnarray}
The most important remark we can draw from this calculation is that the fuzzy sphere phase bifurcates into two distinct phases ( the weak coupling and the strong coupling phases of the gauge field ) almost as soon as we tune on a non-zero mass. The models with and without a mass term are indeed very different.

\begin{figure}
\begin{center}
\includegraphics[width=7cm,angle=-90]{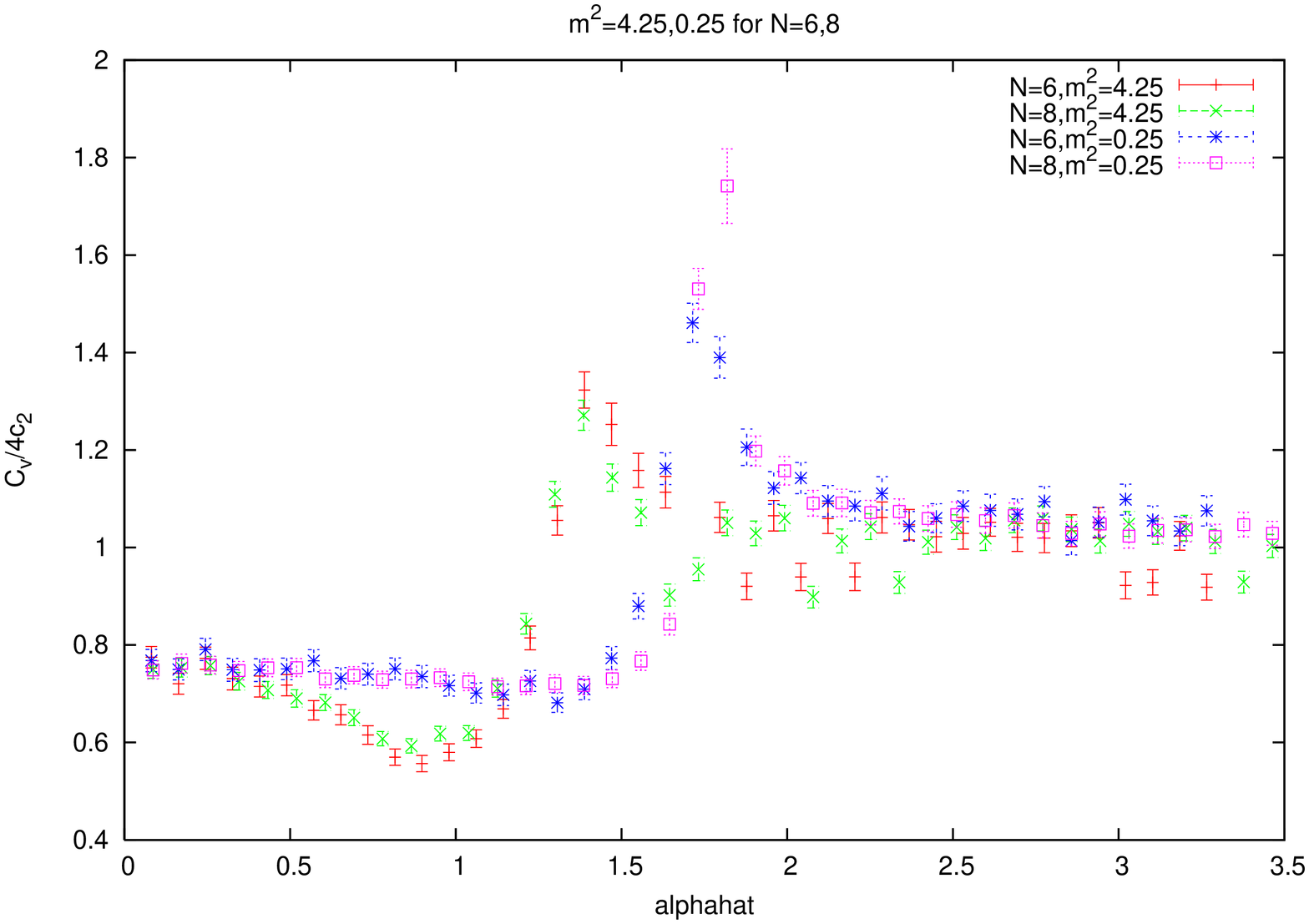}
\includegraphics[width=7cm,angle=-90]{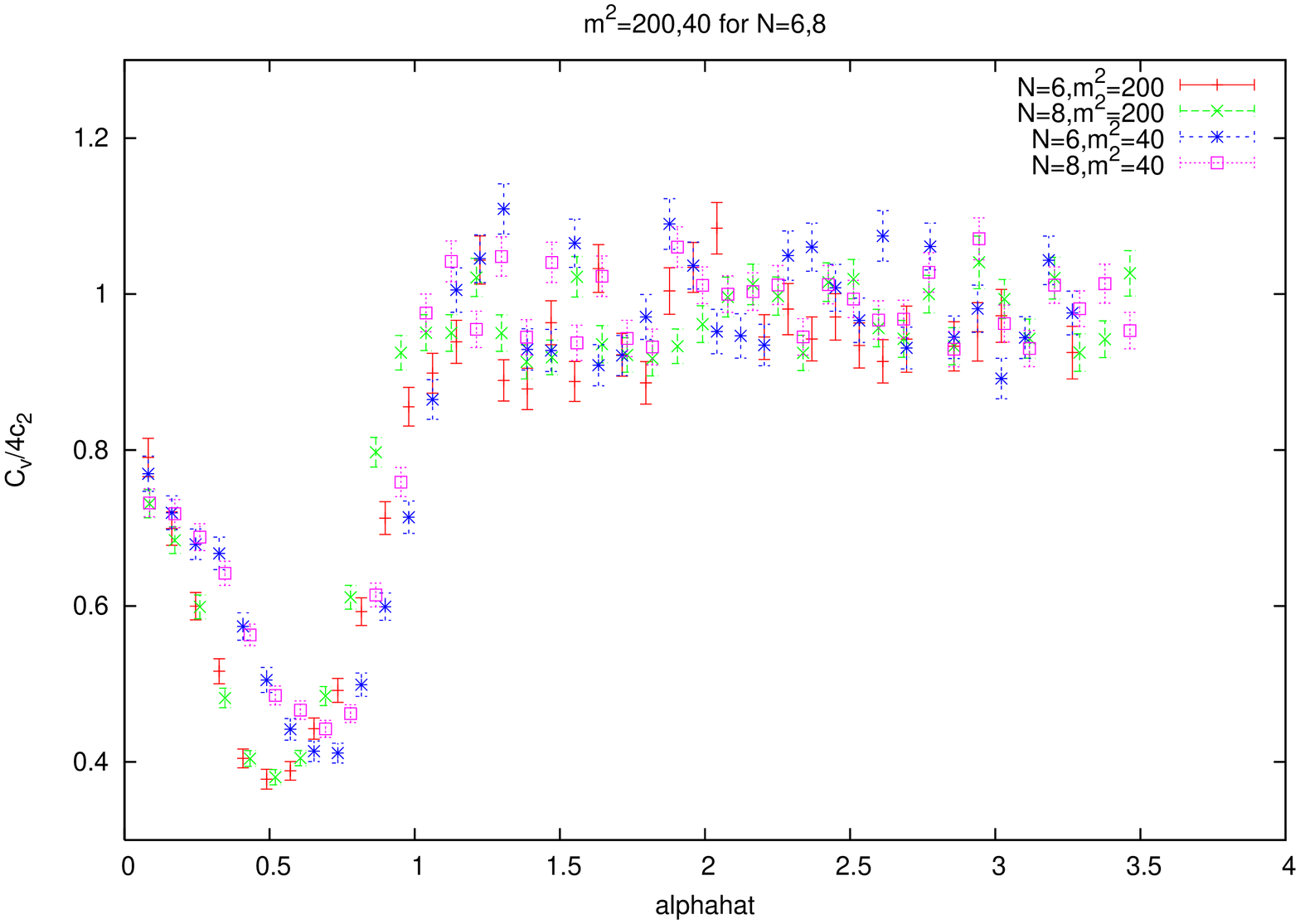}
\caption{{ The specific heat for different values of $m^2$.}}
\end{center}
\end{figure}

\begin{figure}
\begin{center}
\includegraphics[width=10cm,angle=-90]{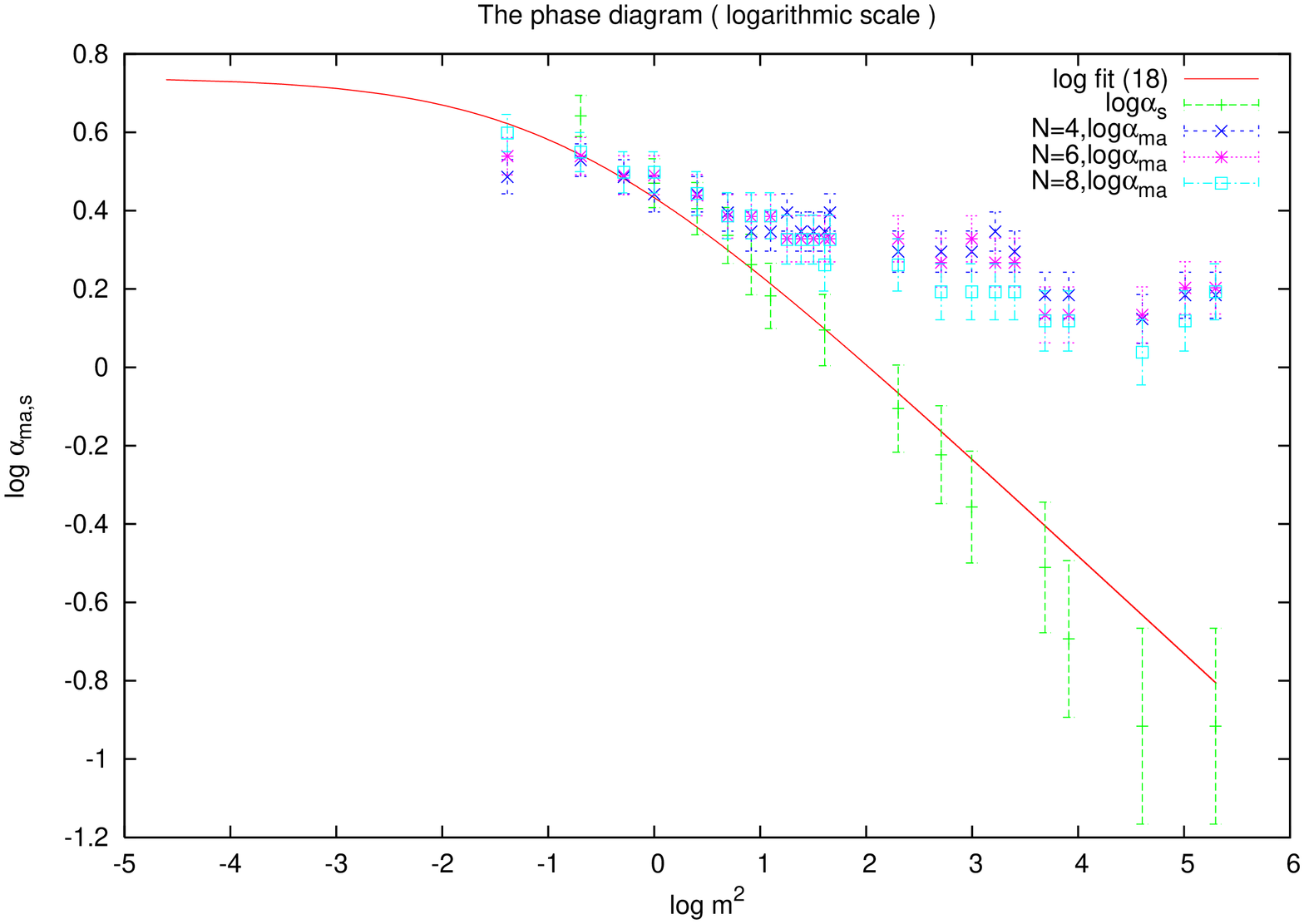}
\includegraphics[width=10cm,angle=-90]{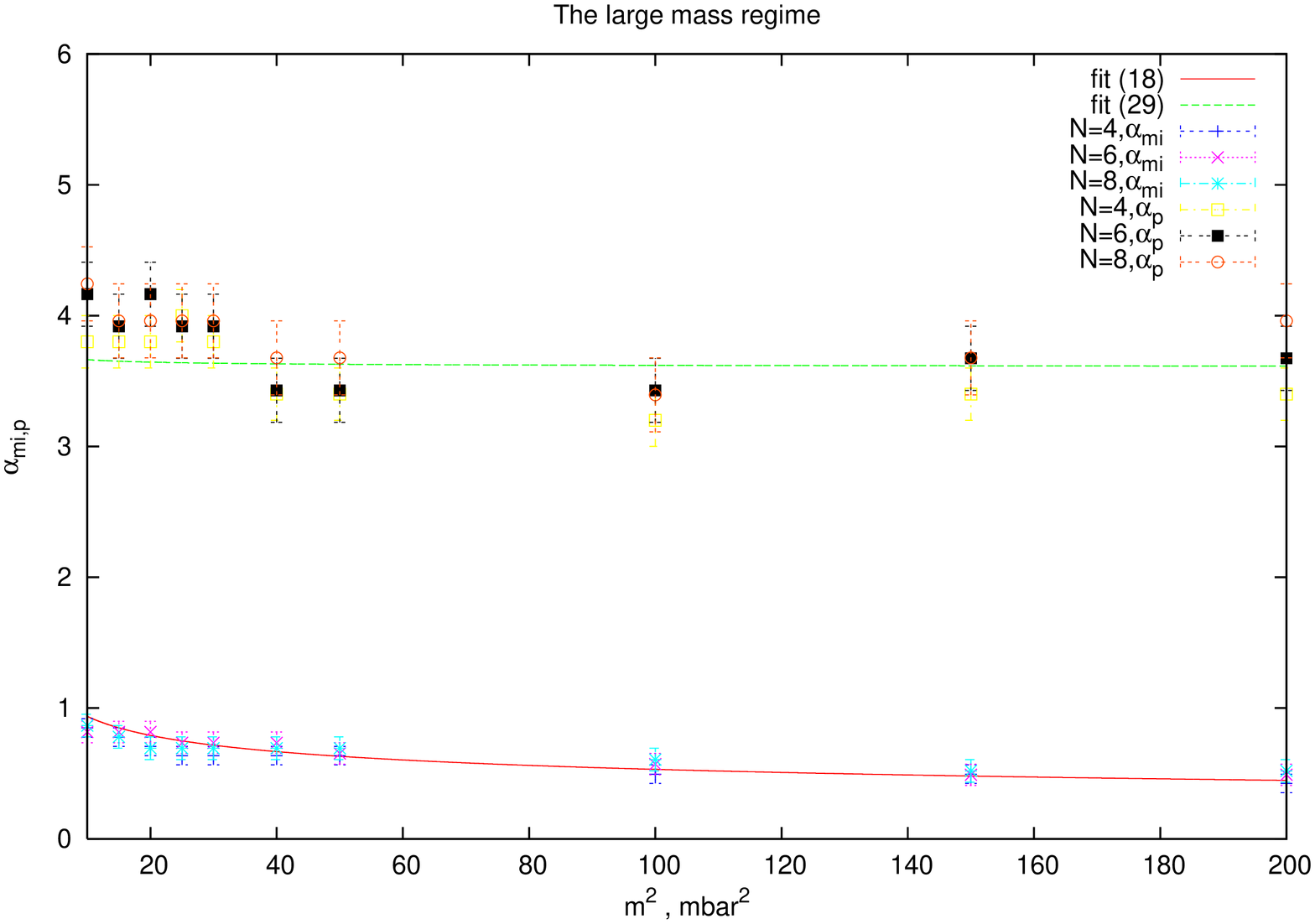}
\caption{{  The phase diagram of the one-plaquette phase transition. For large values of $m$ the ${\alpha}_{p}-$line is the critical one-plaquette transition line while the ${\alpha}_{mi}-$line ( or equivalently the ${\alpha}_{s}-$line ) is the critical fuzzy sphere-to-matrix transition line. For small values of $m$ the ${\alpha}_{ma}-$line coincides with the ${\alpha}_{s}-$line which is the critical fuzzy sphere-to-matrix transition line in this regime. These two lines seem to bifurcate at the triple point.}}
\end{center}
\end{figure}

\section{The one-plaquette model and $1/N$ expansion}

In this section we will follow \cite{ref1}.

\subsection{The one-plaquette variable $W$}

We start by making the observation that in the large $m{\longrightarrow}{\infty}$ limit we can set $\Phi =0$ since the normal scalar field becomes infinitely heavy ( $m$ is precisely its mass ) and thus decouples from the rest of the dynamics. Hence we can effectively impose the extra constraint $X_a^2={\alpha}^2c_2$ on the field $X_a$ in this limit $m{\longrightarrow}{\infty}$. The action (\ref{action}) with $\beta =-2Mc_2{\alpha}^2$  becomes in the limit $m{\longrightarrow}{\infty}$ first and then $N{\longrightarrow}{\infty}$ a commutative $U(1)$ action on the ordinary sphere.

The aim is to relate the action (\ref{action}) with the one-plaquette action. To this end we introduce the $2N{\times}2N$ idempotent 
\begin{eqnarray}
{\gamma}=\frac{1}{N}({\bf 1}_{2N}+2{\sigma}_aL_a)~,~{\gamma}^2=1
\end{eqnarray}
 where ${\sigma}_a$ are the usual Pauli matrices. It has eigenvalues $+1$ and $-1$ with multiplicities $N+1$ and $N-1$ respectively. We introduce the covariant derivative $D_a=L_a+A_a$ through a gauged idempotent ${\gamma}_D$ as follows
\begin{eqnarray}
&&{\gamma}_D=\hat{\gamma}\frac{1}{\sqrt{\hat{\gamma}^2}}\nonumber\\
&&\hat{\gamma}=\frac{1}{N}(1+2{\sigma}_aD_a)={\gamma}+\frac{2}{N}{\sigma}_aA_a~,~~\hat{\gamma}^2=1+\frac{8\sqrt{c_2}}{N^2}\Phi+\frac{2}{N^2}{\epsilon}_{abc}{\sigma}_cF_{ab}.\label{eq0}
\end{eqnarray}
Since we are interested in the large $m{\longrightarrow}{\infty}$ limit we set $\Phi=0$. 
Clearly ${\gamma}_D$ has the same spectrum as ${\gamma}$. In fact  ${\gamma}_D$ is an element of the $d_N-$Grassmannian manifold $U(2N)/U(N+1){\times}U(N-1)$ and hence it contains the correct number of degrees of freedom  $
d_N=4N^2-(N+1)^2-(N-1)^2=2N^2-2$ which is found in a gauge theory on the fuzzy sphere without normal scalar field.

The original $U(N)$ gauge symmetry acts on the covariant derivatives $D_a$ as $D_a^g=g{D}_ag^{+}$, $g{\in}U(N)$. This symmetry  will be enlarged to the following $U(2N)$ symmetry. First we introduce another covariant derivative $D_a^{'}=L_a+A_a^{'}$ through a gauged idempotent ${\gamma}_{D^{'}}$ given by a similar equation to (\ref{eq0}). As before we will also set ${\Phi}^{'}=0$. From the two idempotents ${\gamma}_{D}$ and ${\gamma}_{D^{'}}$ we construct the link variable $W$ as follows
\begin{eqnarray}
W={\gamma}_{D^{'}}{\gamma}_{D}.
\end{eqnarray}
The extended  $U(2N)$ symmetry will then act on $W$ as follows $W{\longrightarrow}VWV^{+}~,~V{\in}U(2N)$. 
This transformation property of $W$ can only be obtained if we  impose the following transformation properties ${\gamma}_{D^{'}}{\longrightarrow}V{\gamma}_{D^{'}}V^{+}$ and ${\gamma}_D{\longrightarrow}V{\gamma}_DV^{+}$ on ${\gamma}_{D^{'}}$ and ${\gamma}_D$ respectively. Hence the $U(N)$ subgroup of this $U(2N)$ symmetry which will act on $D_a$ as $D_a{\longrightarrow}g{D}_ag^{+}$ will also have to act on $D_a^{'}$ as $D_a^{'}{\longrightarrow}g{D}_a^{'}g^{+}$. Under these transformations the gauge fields $A_a$ and $A_a^{'}$ transform as  $A_a{\longrightarrow}gA_ag^++g[L_a,g^+]$ and $A_a^{'}{\longrightarrow}gA_a^{'}g^++g[L_a,g^+]$ respectively like we want. Remark also that for every fixed configuration $A_a^{'}$ the link variable $W$ contains the same degrees of freedom contained in $ {\gamma}_{D}$. 

The main idea is that we want to reparametrize  the gauge field on ${\bf S}^2_N$ in terms of the fuzzy link variable $W$ and the normal scalar field $\Phi$. In other words we want to replace the triplet $(A_1,A_2,A_3)$ with  $(W,\Phi)$ where $W$ is the  object which contains the  degrees of freedom of the gauge field which are tangent to the sphere. Thus in summary we have the coordinate transformation $
(A_1,A_2,A_3){\longrightarrow}(W,\Phi)$. We can check that we have the correct measure, viz
\begin{eqnarray}
\int dA_1dA_2dA_3~{\propto}~\int dWd\Phi. \label{meas}
\end{eqnarray}
It remains now to show that the enlarged $U(2N)$ symmetry reduces to its $U(N)$ subgroup in the large $N$ limit. The starting point is the $2N-$dimensional one-plaquette actions with positive coupling constants ${\lambda}$ and ${\lambda}^{'}$, viz
\begin{eqnarray}
S_P^{}&=&\frac{N}{{\lambda}}Tr_{2N}(W+W^{+}-2)~,~
S_P^{'}=-\frac{N}{{\lambda}^{'}}Tr_{2N}(W^2+W^{+2}-2).\label{plaquette1}
\end{eqnarray}
We have the path integral

\begin{eqnarray}
Z_P~{\propto}~\int d{\gamma}_{D^{'}}d{\Phi}^{'} {\delta}({\Phi}^{'})\int_{W={\gamma}_{D^{'}}{\gamma}_{D}}dWd\Phi {\delta}({\Phi})e^{S_P^{}+S_P^{'}}.\label{path0}
\end{eqnarray}
The extra integrations over ${\gamma}_{D^{'}}$ and ${\Phi}^{'}$ ( in other words over $D^{'}_a$ ) is included in order to maintain gauge invariance of the path integral. The integration over $W$ is done along the orbit $W={\gamma}_{D^{'}}{\gamma}_{D}$ inside the full $U(2N)$ gauge group.  In the large $N$ limit this path integral can be written as
\begin{eqnarray}
Z_P=\int dA^{'}_a{\delta}({\Phi}^{'})\int_{W={\gamma}_{D^{'}}{\gamma}_{D}}dA_a {\delta}({\Phi})e^{S_P+S_P^{'}}.\label{path}
\end{eqnarray}
We need now to check what happens to the actions $S_P^{}$ and $S_P^{'}$ in the large $N$ limit. We introduce the $6$ matrices $2\bar{A}_a=A_a-A_a^{'}$ and $2\hat{A}_a=A_a+A_a^{'}$ with the transformation laws $\bar{A}_a{\longrightarrow}g\bar{A}_ag^+$ and $\hat{A}_a{\longrightarrow}g\hat{A}_ag^++g[L_a,g^+]$. For  the continuum limit of the action $S_P^{}+S_P^{'}$ we obtain after a long calculation the effective theory \footnote{The path integral over the three matrices $\hat{A}_a$ is dominated in the large $N$ limit by the configurations $\hat{A}_a=0$.}\cite{ref1}
\begin{eqnarray}
Z_P^{'}=\int d\bar{A}_a{\delta}\bigg(\frac{1}{2}\{x_a,\bar{A}_a\}\bigg)e^{S_P^{\rm eff}}\label{177}
\end{eqnarray}
where

\begin{eqnarray}
S_P^{\rm eff}&=&N^2\log(\frac{N\pi{\lambda}_1}{8})-\frac{16}{{\lambda}_1 N^3}Tr\bigg(i[L_a,\bar{A}_b]-i[L_b,\bar{A}_a]+{\epsilon}_{abc}\bar{A}_c\bigg)^2+O(\frac{1}{\lambda N^4})-O(\frac{1}{{\lambda}^{'} N^4}).\nonumber\\
\end{eqnarray}
The coupling constant ${\lambda}_1$ ( which is assumed positive in this classical theory for simplicity ) is defined in terms of $\lambda$ abd ${\lambda}^{'}$ by 
\begin{eqnarray}
-\frac{1}{{\lambda}_1}=\frac{1}{{\lambda}}-\frac{4}{{\lambda}^{'}}.
\end{eqnarray}
Notice that this effective action is invariant not only under the trivial original gauge transformation law $\bar{A}_a{\longrightarrow}\bar{A}_a$ but also it is invariant under the non-trivial gauge transformation $\bar{A}_a{\longrightarrow}\bar{A}_a+g[L_a,g^+]$ where $g{\in}U(N)$. This emergent new gauge transformation of $\bar{A}_a$ is identical to the transformation property of a $U(1)$ gauge field on the sphere. Therefore the action $S_P^{\rm eff}$ given by the above equation  is essentially the same $U(1)$ action $-(S-S_0)$ obtained from (\ref{action}) provided we make the following identification 
\begin{eqnarray}
\frac{16}{ N^2{\lambda}_1}\equiv \frac{16}{N^2}(-\frac{1}{{\lambda}}+\frac{4}{{\lambda}^{'}})=\frac{1}{4g^2}\equiv \frac{\tilde{\alpha}^4}{4}\equiv \frac{\bar{\alpha}^4}{4N^2}\label{mmm}
\end{eqnarray}
between the $U(1)$ gauge coupling constant $g$ on the fuzzy sphere and the one-plaquette model coupling constant ${\lambda}_1$. Let us also remark that in this large $N$ limit in which $g$ is kept fixed the one-plaquette coupling constant ${\lambda}_1 $ goes to zero. Hence the fuzzy sphere action with fixed coupling constant $g$ corresponds in this particular limit to the one-plaquette gauge field in the weak regime and agreement between the two is expected only for weak couplings ( large values of $\tilde{\alpha}$ ).

\subsection{The one-plaquette path integral}
Let us decompose the $2N{\times}2N$ matrix $W$ as follows 
\begin{eqnarray}
W=\left(\begin{array}{cc}
{W}_1 & {W}_{12}\\
 -{W}_{12}^+& -{W}_2
\end{array}
\right). 
\end{eqnarray}
 In particular ${W}_1={W}_1^+$ is an $(N+1)\times (N+1)$ matrix, ${W}_2={W}_2^+$ is an $(N-1)\times (N-1)$ matrix and ${W}_{12}$ is an $(N+1)\times (N-1)$ matrix whereas the hermitian adjoint ${W}_{12}^+$ is an $(N-1)\times (N+1)$ matrix. Since $W^+W=1$ we have the conditions 
\begin{eqnarray}
&&{W}_1^+W_1+{W}_{12}{W}_{12}^+=1~,~{W}_2^+W_2+{W}_{12}^+{W}_{12}=1~,~{W}_1{W}_{12}+{W}_{12}{W}_2=0.\label{decomp10}
\end{eqnarray}
 Let us recall that since the integration over $W$ is done along the orbit $W={\gamma}_{D^{'}}{\gamma}_{D}$ inside $U(2N)$ and since in the large $N$ limit both ${\gamma}_{D^{'}}$ and ${\gamma}_{D}$ approach the usual chirality operator $\gamma =n_a{\sigma}_a$ we see that $W$ approaches the identity matrix in this limit. Thus we have the behaviour $W_1=({\gamma}_{D^{'}}{\gamma}_{D})_1{\longrightarrow}{\bf 1}_{N+1}$, $-W_2=-({\gamma}_{D^{'}}{\gamma}_{D})_2{\longrightarrow}{\bf 1}_{N-1}$ and $W_{12}=({\gamma}_{D^{'}}{\gamma}_{D})_{12}{\longrightarrow}0$.

The main approximation adopted in \cite{ref1} consisted in replacing  the constraint $W={\gamma}_{D^{'}}{\gamma}_{D}$ with the simpler constraint $W{\longrightarrow}{\bf 1}_{2N}$ by taking the diagonal parts $W_1$ and $-W_2$  to be {\it two arbitrary}, i.e independent of ${\gamma}_{D^{'}}$, unitary matrices which are very close to the identities ${\bf 1}_{N+1}$ and ${\bf 1}_{N-1}$ respectively while allowing the off-diagonal parts  $W_{12}$ and $W_{12}^{+}$ to go to zero. We observe that by including only $W_1$ and $-W_2$  in this approximation we are including  in the limit precisely the correct number of degrees of freedom tangent to the sphere, viz $2N^2$. Thus in this approximation the integrations over $\Phi$, ${\Phi}^{'}$ and ${\gamma}_{D^{'}}$ decouple while the integrations over $W_{12}$ and $W_{12}^+$ are dominated by $W_{12}=W_{12}^+=0$. There remains the two independent path integrals over $W_1$ and $-W_2$  which  are clearly equal  in the strict limit since the matrix dimension of $W_1$ approaches  the matrix dimension of $-W_2$ for large $N$. Thus the path integral $Z_P^{'}$ reduces  to

\begin{eqnarray}
Z_P^{'}&{\propto}&[Z_P(\lambda,{\lambda}^{'})]^2\label{188}
\end{eqnarray}
where
\begin{eqnarray}
Z_P(\lambda,{\lambda}^{'})&=&\int dW_1 \exp\bigg\{\frac{N}{\lambda}Tr(W_1+W_1^+-2)-\frac{N}{{\lambda}^{'}}Tr(W_1^2+W_1^{+2}-2)\bigg\}.\label{189}
\end{eqnarray}

\subsection{Saddle point solution}

The path integral of  a $2-$dimensional $U(N)$ gauge theory in the axial gauge $A_1=0$  on a lattice with volume $V$ and lattice spacing $a$ is given by $Z_P(\lambda,\infty)^{V/a^2}$ where $Z_P(\lambda,\infty)$ is the above partition function (\ref{189}) for ${\lambda}^{'}=\infty $, i.e the partition function of the one-plaquette model $S_P=\frac{N}{{\lambda}}Tr(W_1+W_1^{+}-2)$. Formally the partition function $Z_P(\lambda,{\lambda}^{'})^{V/a^2}$ for any value of the coupling constant ${\lambda}^{'}$ can be obtained by expanding the model $S_1+S_1^{'}$ around ${\lambda}^{'}=\infty $. Thus it is not difficult to observe that the one-plaquette action  $S_P+S_P^{'}$ does lead to  a more complicated  $U(N)$ gauge theory in two dimensions. 

Therefore we can see that  the partition function $Z_P^{'}$ of a $U(1)$ gauge field on the fuzzy sphere  is proportional to the partition function of  a {\it generalized} $2-$dimensional $U(N)$ gauge theory in the axial gauge  $A_1=0$  on a lattice with two plaquettes. This doubling of plaquettes is reminiscent of the usual doubling of points in Connes standard model. We are therefore interested in the $N-$dimensional one-plaquette model
\begin{eqnarray}
Z_P(\lambda,{\lambda}^{'})&=&\int dW exp\bigg(\frac{N}{\lambda}Tr(W+W^+-2)-\frac{N}{{\lambda}^{'}}Tr(W^2+W^{+2}-2)\bigg).\label{193}
\end{eqnarray}
Let us recall that $dW$ is the $U(N)$ Haar measure. We can immediately diagonalize the link variable $W$ by writing $W=TDT^{+}$ where $T$ is some $U(N)$ matrix and $D$ is diagonal with elements equal to the eigenvalues $exp(i{\theta}_i)$ of $W$. In other words $D_{ij}={\delta}_{ij}exp(i{\theta}_i)$. The integration over $T$ can be done trivially and one ends up with the path integral
\begin{eqnarray}
Z_P(\lambda,{\lambda}^{'})=\int {\prod}_{i=1}^{N}d{\theta}_i e^{NS_N}.\label{194}
\end{eqnarray}
The action $S_N$ is given by

\begin{eqnarray}
S_N&=& \frac{2}{\lambda}\sum_{i} \cos{\theta}_i-\frac{2}{{\lambda}^{'}}\sum_{i} \cos2{\theta}_i+\frac{1}{2N}\sum_{i{\neq}j}\ln \bigg(\rm sin\frac{{\theta}_i-{\theta}_j}{2}\bigg)^2-\frac{2N}{\lambda}+\frac{2N}{{\lambda}^{'}}.\label{1p}
\end{eqnarray}
Since the link variable $W$ tends to one in the large $N{\longrightarrow}\infty$ limit we can conclude that all the angles ${\theta}_i$ tend to $0$ in this limit and thus we can   consider instead of the full one-plaquette model action 
(\ref{1p}) the {\it small} one-plaquette model action 
\begin{eqnarray}
S_N
&=&-\frac{1}{{\lambda}_2}\sum_i {\theta}_i^2+\frac{1}{2N}\sum_{i{\neq}j}\ln \frac{\big({\theta}_i-{\theta}_j\big)^2}{4}+O({\theta}^4).
\label{exx2}
\end{eqnarray}
${\lambda}_2$ is given by 
\begin{eqnarray}
&&\frac{2}{{\lambda}_2}=-\frac{2}{{\lambda}_1}+\frac{1}{6}.\label{lambda2}
\end{eqnarray}
For the consistency of the solution below the coupling constant ${\lambda}_1$ must be negative ( as opposed to the classical model  where ${\lambda}_1$ was assumed positive ) and as a consequence the coupling constant ${\lambda}_2$ is always positive. As it turns out most of the classical arguments of section $4.1$ will go through unchanged when  ${\lambda}_1$  is taken negative. Thus in this present quantum theory of the model we will identify the effective one-plaquette action $S_P^{\rm eff}$ with the fuzzy sphere action 
$S-S_0$ ( which is to be compared with the classical identification $-S_P^{\rm eff}=S-S_0$) and hence we must make the following identification of the coupling constants
\begin{eqnarray}
-\frac{16}{ N^2{\lambda}_1}= \frac{1}{4g^2}=\frac{\bar{\alpha}^4}{4N^2}.
\end{eqnarray}
Furthermore it is quite obvious that the expansion (\ref{exx2}) will only be valid for small angles ${\theta}_i$ in the range $
-\frac{1}{2}{\leq}{\theta}_i{\leq}\frac{1}{2}$. Let us also note that the action (\ref{exx2}) can be obtained from the effective one-plaquette model
\begin{eqnarray}
S_P^{\rm eff}&=& \frac{2}{{\lambda}_2^{\rm eff}}Tr(W_{\rm eff}+W_{\rm eff}^{+}-2)\nonumber\\
&=&\frac{2}{{\lambda}_2^{\rm eff}}\sum_{i} \cos{\theta}_i^{\rm eff}-\frac{2N}{{\lambda}_2^{\rm eff}}.\label{lrim}
\end{eqnarray}
For small ${\theta}_i^{\rm eff}$ in the range $
-{1}{\leq}{\theta}_i^{\rm eff}{\leq}{1} $ the total effective one-plaquette action becomes
\begin{eqnarray}
S_N^{\rm eff}&=& 
-\frac{1}{{\lambda}_2^{\rm eff}}\sum_{i} ({\theta}_i^{{\rm eff}})^2+\frac{1}{2N}\sum_{i{\neq}j}\ln \frac{\big({\theta}_i^{\rm eff}-{\theta}_j^{\rm eff}\big)^2}{4}+O(({\theta}^{\rm eff})^4).\label{lj}
\end{eqnarray}
The action (\ref{lj}) must be  identical to the action (\ref{exx2}) and hence we must have ${\theta}_i^{\rm eff}=2{\theta}_i$ and ${\lambda}_2^{\rm eff}=4{\lambda}_2$.
 
The saddle point solution of the action (\ref{lrim}) must satisfy the equation of motion
\begin{eqnarray}
\frac{2}{{\lambda}_2^{\rm eff}}\sin{\theta}_i^{\rm eff}=\frac{1}{N}\sum_{j{\neq}i}\cot\frac{{\theta}_i^{\rm eff}-{\theta}_j^{\rm eff}}{2}.\label{eomla}
\end{eqnarray}
In the continuum  large $N$ limit we introduce a density of eigenvalues ${\rho}(\theta)$ and the equation of motion becomes

\begin{eqnarray}
\frac{2}{{\lambda}_2^{\rm eff}}\sin{\theta}_{\rm eff}=\int_{}^{}d{\tau}_{\rm eff} {\rho}({\tau}_{\rm eff})\cot\frac{{\theta}_{\rm eff}-{\tau}_{\rm eff}}{2}.
\end{eqnarray}
By using the expansion $\cot\frac{{\theta}-{\tau}}{2}=2\sum_{n=1}^{\infty}\big(\sin n\theta \cos n\tau -\cos n\theta \sin n\tau \big)$ we can solve this equation quite easily in the strong-coupling phase ( large values of ${\lambda}_2$ ) and one finds the solution
\begin{eqnarray}
{\rho}({\theta}_{\rm eff})=\frac{1}{2{\pi}}+\frac{1}{\pi {\lambda}_2^{\rm eff}} \cos {\theta}_{\rm eff}.\label{198}
\end{eqnarray}
However it is obvious that this solution makes sense only  where the density of eigenvalues is positive definite, i.e for ${\lambda}_2^{\rm eff}$ such that
\begin{eqnarray}
\frac{1}{2\pi}-\frac{1}{\pi{\lambda}_2^{\rm eff}}{\geq}0~{\Leftrightarrow}({\lambda}_2^{\rm eff})^*=2~{\Leftrightarrow}{\lambda}_2^*=0.5.\label{199}
\end{eqnarray}
In the continuum large $N$ limit where $ \tilde{\alpha}^4$ is kept fixed instead of ${\lambda}_1$ we can see that $\frac{1}{{\lambda}_1}$ scales with $N^2$ and as a consequence ${\lambda}_2=-{{\lambda}_1}=\frac{64}{N^2\tilde{\alpha}^4}$. Thus the critical value ${{\lambda}_2}^*=0.5$ leads to the critical value 
\begin{eqnarray}
\bar{\alpha}_*^4=\frac{64}{{\lambda}_2^*}=128~{\Leftrightarrow}~\bar{\alpha}_*=3.36
\end{eqnarray}
 which is to be compared with the observed value 
\begin{eqnarray}
\bar{\alpha}_*=3.35{\pm}0.25.
\end{eqnarray}
This strong-coupling solution (\ref{198}) should certainly work for large enough values of ${\lambda}_2$. However this is not the regime we want. To find the solution for small values of ${\lambda}_{2}$ the only difference  with the above analysis is that the range of the eigenvalues is now $[-{\theta}_*,+{\theta}_*]$ instead of $[-\pi,+\pi]$ where ${\theta}_*$ is an angle less than $\pi$ which is a function of ${\lambda}_{2}$. It is only in this regime of small ${\lambda}_{2}$ where the fuzzy sphere action with fixed coupling constant $g$ is expected to correspond to the one-plaquette model. Indeed the fact that $W{\longrightarrow}1$ in the large $N$ limit is equivalent to the statement that the one-plaquette model is in the very weak-coupling regime.  In the strong-coupling region deviations become significant near the sphere-to-matrix transition point.

In the 'very' weak-coupling regime the saddle point equation reduces to
\begin{eqnarray}
\frac{2{\theta}_i}{{\lambda}_2}=\frac{2}{N}\sum_{j{\neq}i}\frac{1}{{\theta}_i-{\theta}_j}
\label{erlm2}
\end{eqnarray}
This problem was easily solved using matrix theory techniques in \cite{ref1}. See also \cite{gross}. In the large $N{\longrightarrow}\infty $ we find the density of eigenvalues 
\begin{eqnarray}
&&{\rho}(\theta)=\frac{1}{\pi{\lambda}_2}\sqrt{2{\lambda}_2-{\theta}^2}.\label{rho}
\end{eqnarray}
It is obvious that this density of eigenvalues is only defined for angles $\theta$ which are in the range $-\sqrt{2{\lambda}_2} {\leq}\theta {\leq}\sqrt{2{\lambda}_2}$. However the value of the critical angle ${\theta}_*$ should be determined from the normalization condition $\int_{-{\theta}_*}^{{\theta}_*} d\theta {\rho}(\theta)=1$. This condition yields the value
\begin{eqnarray}
{\theta}_*=\sqrt{2{\lambda}_2}.
\end{eqnarray}
The fuzzy one-plaquette third order phase transition happens at the value of the coupling constant  ${\lambda}_2=0.5$ where the eigenvalues $e^{i{{\theta}_i}}$ fill half of the unit circles. This half is due to the fact that ${\theta}_i^{\rm eff}=2{\theta}_i$.

In \cite{ref1} we also computed the predictions coming from this model for  the free energy and specific heat. We found very good agreement between the fuzzy one-plaquette model and the data in the weak-coupling phase and even across the transition point to the strong-coupling phase until the matrix-to-sphere transition point where deviations become significant. In particular the specific heat is found to be equal to $1$ in the fuzzy sphere-weak coupling phase of the gauge field which agrees with the observed value $1$ seen in our Monte Carlo simulation. The value $1$ comes precisely because we have two plaquettes which approximate the noncommutative $U(1)$ gauge  field on the fuzzy sphere.

\section{Conclusion}
In this article we have determined to a large extent the phase diagram of noncommutative $U(1)$ gauge theory in two dimensions using the fuzzy sphere as a non-perturbative regulator.  The central tool we employed was  Monte Carlo simulation and in particular the Metropolis algorithm. 

We have identified three distinct phases. $1)$ A matrix phase in the strong coupling regime. The order parameter $<TrX_aX_a>=0$ in this phase. $2)$ A fuzzy sphere phase at weak coupling with order parameter $<TrX_aX_a>{\neq}0$ and with constant specific heat.  $3)$ A new strong coupling fuzzy sphere phase. Here the fluctuations are around a fuzzy sphere background, i.e $<TrX_aX_a>{\neq}0$, in addition the specific heat is non-constant in this phase. 

The transition between the weak and strong coupling fuzzy sphere phases is third order. The other two transitions appear to be first order. We have clear numerical evidence for a jump in the internal energy $<S>$ between the matrix and weak coupling fuzzy sphere phase. The corresponding jump in $<S>$ has become smaller or disappeared in the strong coupling fuzzy sphere to matrix phase transition. However the order parameter $<TrX_aX_a>$ still jumps discontinuously. We observe that for the $m=0$ model the specific heat becomes constant in both the strong coupling matrix phase ( $Cv=\frac{3}{4}N^2$ ) and the weak coupling fuzzy sphere phase  ( $Cv=N^2$ ). As the mass $m^2$ increases a new third phase opens up and  the three phases meet at a triple point.

In this article we have also confirmed the theoretical one-loop prediction of the ${\bf S}^2_N-$to-matrix critical line \cite{ref}. The transition between strong and weak couplings fuzzy sphere phases is found to agree with the $\frac{1}{N}$ expansion prediction of the one-plaquette critical line in the infinite mass limit. It seems that near these lines these approximations ( the one-loop and the one-plaquette ) are essentially exact. We also gave a Monte Carlo measurement of the triple point where the three phases meet.

We would like to indicate that a high precision measurement of the one-plaqutte critical line and the triple point would be highly desirable. We also lack a theoretical control of the triple point. Improvement of the one-plaquette approximation of the NC $U(1)$ gauge field on the fuzzy sphere ${\bf S}^2_N$ is necessary. In particular it would be very interesting to have an alternative more rigorous derivation of the one-plaquette critical value $3.35$. Furthermore we believe that an extension of this approximation to $1)$ the regime of small masses and $2)$ the strong-coupling phase of the gauge theory is possible and needed. The Monte Carlo measurement and the one-loop theoretical description of the ${\bf S}^2_N-$to-matrix critical line are on the other hand very satisfactory.

The most natural generalization of this work is Monte Carlo simulation of fuzzy fermions in two dimensions \cite{15} and fuzzy topological excitations \cite{14}. In particular our current project consists of the simulation of the NC Schwinger model and the NC two dimensional QCD on the fuzzy sphere. Then one must contemplate going to NC $4$ dimensions with full QCD. Early steps towards these goals were taken in \cite{16} and in the first reference of \cite{S2S2}.  Supersymmetric models are also possible and in some sense natural \cite{Bal}. It would be nice to have Monte Carlo control over such supersymmetry.

\paragraph{Acknowledgements}
The author Badis Ydri  would like to thank The Department of Mathematical Physics, NUI Maynooth, Ireland, where a major part of this work was carried out during the spring of 2005. The author Badis Ydri  would also like to thank F.Garcia-Flores for his gracious help with the simulation in the early stages of this research.

\appendix

\section{The order parameters and probability distribution}

In most of this section we restrict our discussion to the case $m=0$. The case $m{\neq}0$ has the same order parameters and probability distribution and we can show that they behave in exactly the same way.

\subsection{Order parameters}

The model (\ref{action}) is symmetric under $U(N)$ gauge transformations of the matrices $X_a$ and as a consequence we can only attach a physical meaning to gauge invariant quantities which are constructed out of $X_a$.  In other words we have to measure gauge invariant observables. Let us introduce the scalar field $\tilde{\Phi}$ defined by
\begin{eqnarray}
&&\tilde{\Phi}=\sqrt{4c_2}{\alpha}^2\Phi+{\alpha}^2c_2{\equiv}\sum_{a=1}^3X_a^2.\label{scalar}
\end{eqnarray}
This field $\tilde{\Phi}$ can be decomposed in the basis of $N{\times}N$ polarization tensors $\hat{Y}_{lm}$ as follows
\begin{eqnarray}
\tilde{\Phi}=\sum_{l=0}^{N-1}\sum_{m=-l}^l{\phi}_{lm}\hat{Y}_{lm}.\label{lm}
\end{eqnarray}
We remark that $\hat{Y}_{00}={\bf 1}_N$, $\hat{Y}_{1 \pm 1}=\frac{\pm 1}{\sqrt{2}}\sqrt{\frac{3}{c_2}}L_{\pm}$, $\hat{Y}_{10}=\sqrt{\frac{3}{c_2}}L_3$ and since $X^{+}=X$ we must also have $({\phi}^{*})_{lm}=(-1)^{m}{\phi}_{l-m}$. The total power in this field is given by
\begin{eqnarray}
&&P{\equiv}<\frac{1}{N}Tr\tilde{\Phi}^2>=<\sum_{l=0}^{N-1}\sum_{m=-l}^l|{\phi}_{lm}|^2 >.
\end{eqnarray}
Another gauge invariant quantity we can measure is the power in the $l=0$  modes defined  by
\begin{eqnarray}
&&P_{0}{\equiv}<\big(\frac{1}{N}Tr\tilde{\Phi}\big)^2>=< {\phi}_{00}^2 >.
\end{eqnarray}
The data for $P$ and $P_{0}$  is given in figure $12$. The collapsed data is given in terms of $\hat{P}=\frac{N^2P}{c_2^2}$ and $\hat{P}_0=\frac{N^2P_{0}}{c_2^2}$ as  functions of $\tilde{\alpha}$. From these results we can conclude that in the fuzzy sphere phase $P=P_0$ and thus  the scalar field $\tilde{\Phi}$ is proportional to the identity matrix since all its power is localized in the zero mode, i.e we have $\tilde{\Phi}=\sum_{a=1}^3X_a^2={\phi}_{00}{\bf 1}_N$. Furthermore  a fit is given by $P=P_0={\alpha}^4c_2^2$ and hence we have  essentially ${\phi}_{00}={\alpha}^2c_2$ in this phase which is consistent with the equilibrium configuration $X_a=\alpha L_a$ as expected. 

This result is confirmed by measuring the observables
\begin{eqnarray}
&&p_1=<\frac{1}{N}TrX_1^2>~,~{\rm etc}.
\end{eqnarray}
The data for $N=6,8$ is shown on figure $13$. The collapsed quantities are   $\hat{p}_1=\frac{Np_1}{c_2}$, etc. We find that in the fuzzy sphere phase we can fit the data to $p_a=\frac{{\alpha}^2c_2}{3}$ which is consistent with the number $\frac{1}{N}TrL_1^2=\frac{1}{N}TrL_2^2=\frac{1}{N}TrL_3^2=\frac{c_2}{3}$.
\begin{figure}[h]
\begin{center}
\includegraphics[width=7cm,angle=-90]{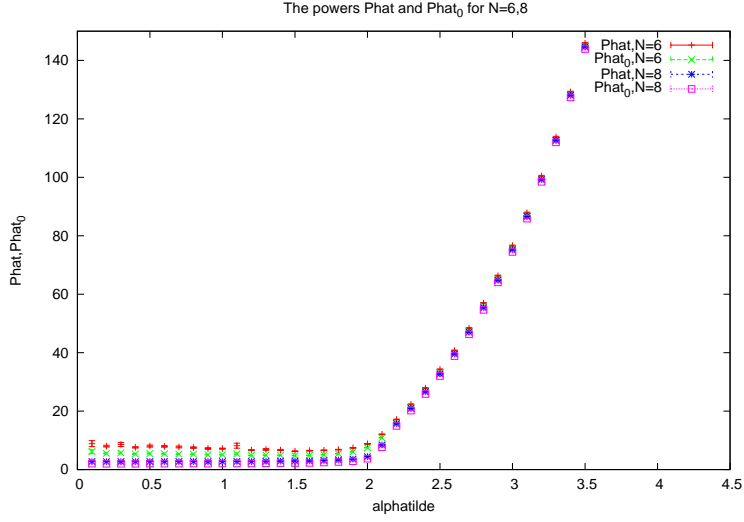}
\caption{{ The powers $\hat{P}$, $\hat{P}_0$ for $N=6,8$.}}
\end{center}
\end{figure}
\begin{figure}[h]
\begin{center}
\includegraphics[width=7cm,angle=-90]{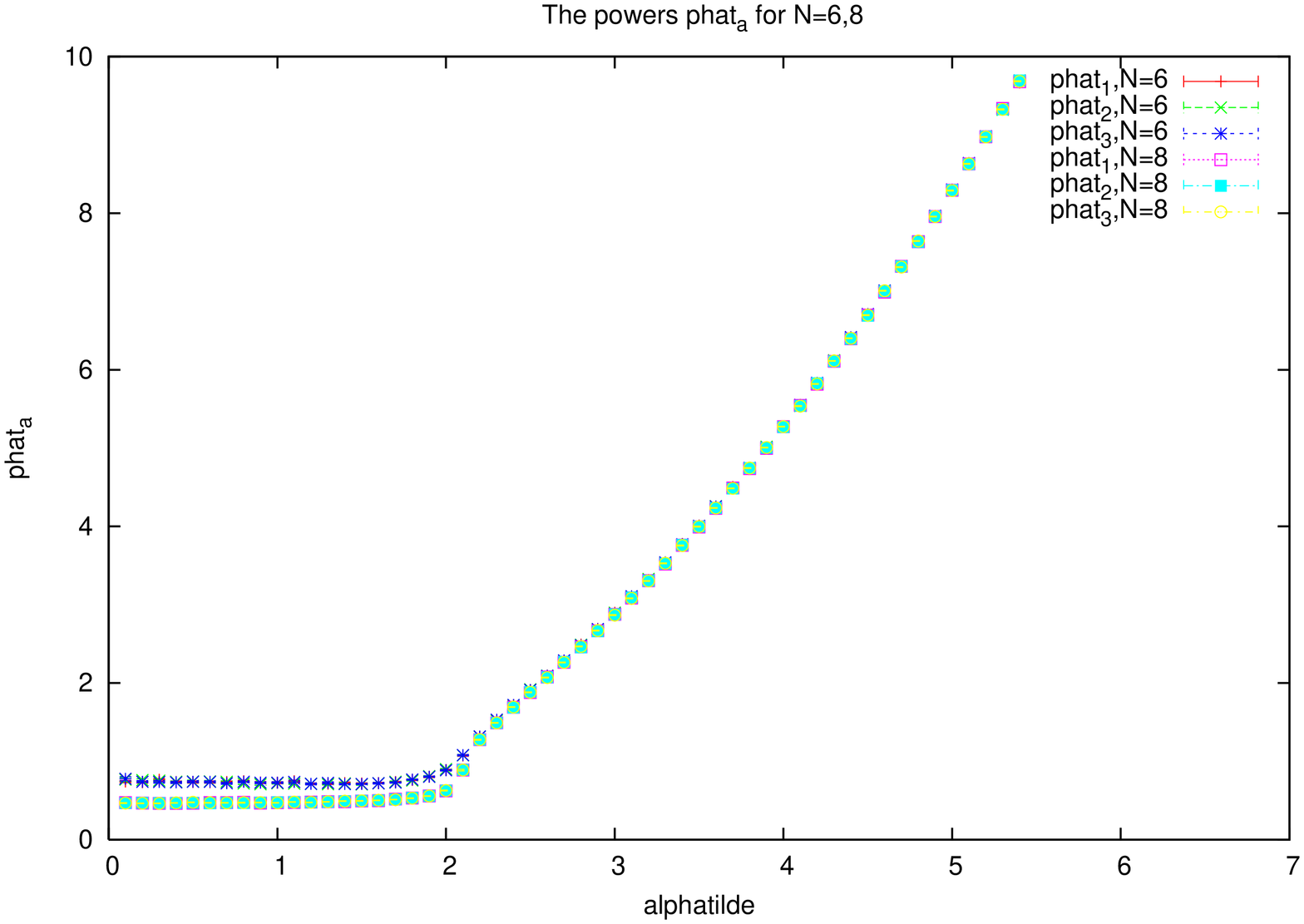}
\caption{{ The powers $\hat{p}_a$ for $N=6,8$.}}
\end{center}
\end{figure}

Let us now introduce the following $3$ scalar fields ($(a,b,c)=(1,2,3),(3,1,2),(2,3,1)$ )
\begin{eqnarray}
&&\Phi_{ab}{\equiv}i[X_a,X_b]
\end{eqnarray}
The total powers associated with these scalar fields are given by
\begin{eqnarray}
&&P_{ab}{\equiv}<\frac{1}{N}Tr\Phi_{ab}^2>=<\sum_{l=0}^{N-1}\sum_{m=-l}^l|(\phi_{ab})_{lm}|^2 >.
\end{eqnarray}
In this case the powers in the $l=0$  modes vanish by construction.
The definition of the modes $({\phi}_{ab})_{lm}$ is obvious by analogy with equation (\ref{lm}). The results are displayed on figure $14$. The collapsed quantities are $\hat{P}_{ab}=\frac{N^2P_{ab}}{c_2}$ . Again we can fit the data to the theoretical prediction $P_{ab}=\frac{{\alpha}^4c_2}{3}$ to a high degree of accuracy in the fuzzy sphere phase. Remark that the Yang-Mills action is given by $<YM>=\frac{N^2}{2}{}\big(P_{12}+P_{31}+P_{23}\big)$. In the fuzzy sphere we clearly have $<YM>=\frac{3N^2}{2}P_{12}
=\frac{3N^2}{2}P_{31}=\frac{3N^2}{2}P_{23}$.

\begin{figure}[h]
\begin{center}
\includegraphics[width=7cm,angle=-90]{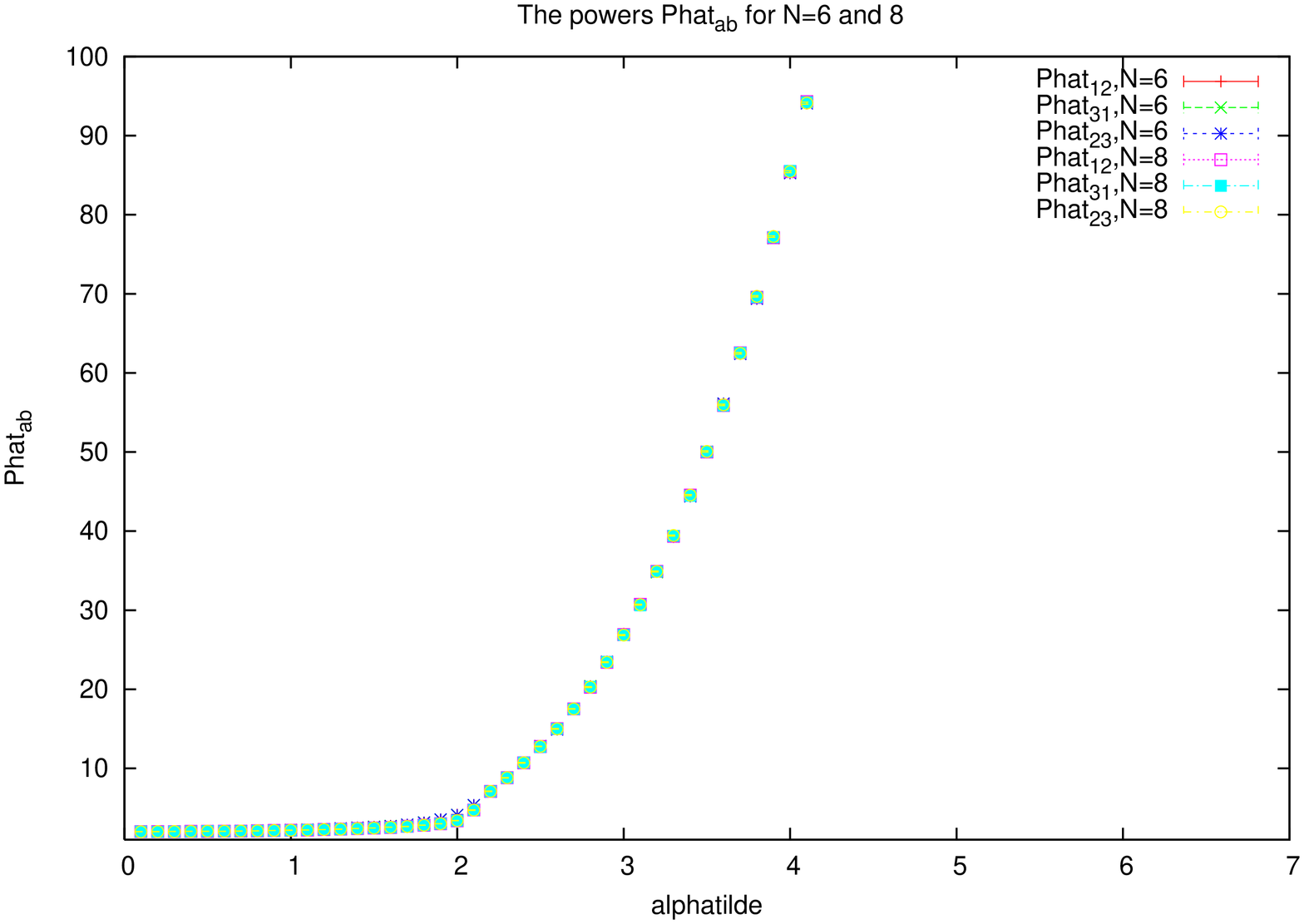}
\caption{{ The powers $\hat{P}_{ab}$ for $N=6,8$.}}
\end{center}
\end{figure}

We will also measure the following gauge invariant quantities 
\begin{eqnarray}
&&p_{cs1}=-i<\frac{1}{N}TrX_1X_2X_3>~,~p_{cs2}=-i<\frac{1}{N}TrX_1X_3X_2>.
\end{eqnarray}
The results are shown on figure $15$. In the fuzzy sphere phase we expect that the power $p_{cs1}$ behaves as $p_{cs1}=\frac{{\alpha}^3c_2}{6}$ whereas the power $p_{cs2}$ behaves as $p_{cs2}=-\frac{{\alpha}^3c_2}{6}$. These are  precisely the correct fits in the fuzzy sphere phase found for $N=4,6$ and $8$ respectively.  The collapsed powers are $\hat{p}_{cs1}=\frac{\sqrt{N^3}p_{cs1}}{4c_2}$, etc. We remark that the Chern-Simons-like action is given by $<CS>=-2{{\alpha}N^2}\big(p_{cs1}-p_{cs2}\big)$. In the fuzzy sphere phase we clearly have $<CS>=-4{{\alpha}N^2}p_{cs1}=4{{\alpha}N^2}p_{cs2}$.
\begin{figure}[h]
\begin{center}
\includegraphics[width=7cm,angle=-90]{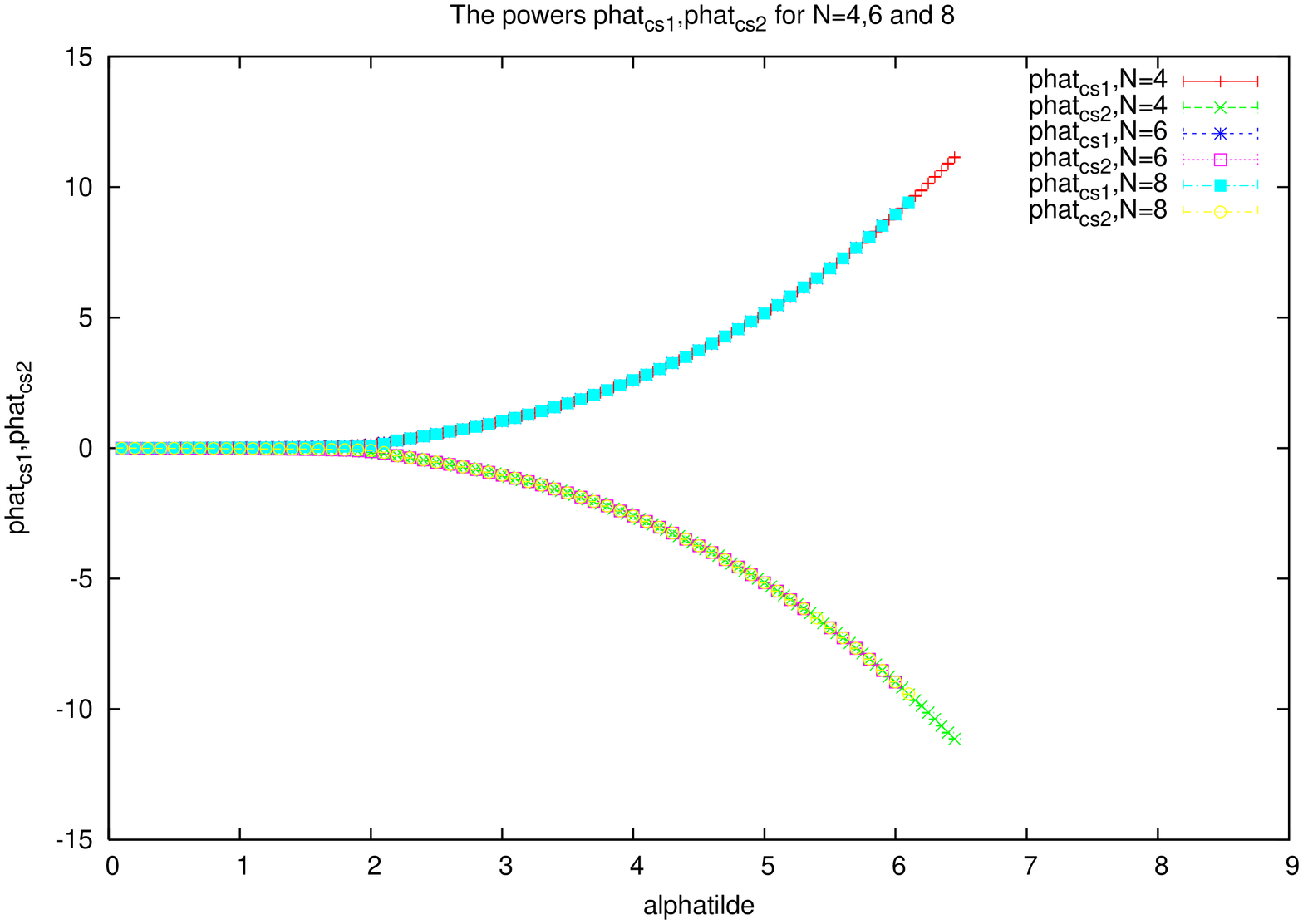}
\caption{{ The operators $\hat{p}_{cs1}$ and $\hat{p}_{cs2}$ for $N=4,6,8$.}}
\end{center}
\end{figure}
\subsection{Geometric interpretation}
The covariant derivatives $X_a$ can in general be expanded in terms of $N{\times}N$ spherical harmonics $\hat{Y}_{lm}$ as follows
\begin{eqnarray}
X_a=\alpha \sum_{b=1}^3x_a^{b}L_b+\bar{X}_a~,~\bar{X}_a=(x_a)_{00}\hat{Y}_{00}+\sum_{l=2}^{\infty}\sum_{m=-l}^{l}(x_a)_{lm}\hat{Y}_{lm}.\label{xbar}
\end{eqnarray}
The three vectors $\vec{x}_a$ are the modes of $X_a$ with angular momentum $l=1$ since $\hat{Y}_{1{\pm}1}={\pm}\sqrt{\frac{{3}}{{2c_2}}}L_{\pm}={\pm}\sqrt{\frac{{3}}{{2c_2}}}(L_1{\pm}iL_2)$ and $\hat{Y}_{10}=\sqrt{\frac{3}{c_2}}L_3$. They are vectors in ${\bf R}^3$ which define the geometry of a  parallelepiped.This geometry is precisely determined by the dynamics of $X_a$ given by the action (\ref{action}). The order parameters $\hat{p}_a$, $\hat{P}_{ab}$ and $\hat{p}_{cs1}-\hat{p}_{cs2}$ have the simple interpretation of the lenghts squared, the areas of the faces squared and the volume respectively of this parallelepiped. We have ( by setting $\bar{X}_a=0$ ) the following expressions
\begin{eqnarray}
&&l^2=\hat{p}_1=\frac{1}{3}\tilde{\alpha}^2\vec{x}_1^2~,~{\rm etc}\nonumber\\
&&a^2=\hat{P}_{12}=\frac{1}{3}\tilde{\alpha}^4(\vec{x}_1{\times}\vec{x}_2)^2~,~{\rm etc}\nonumber\\
&&v=\hat{p}_{cs1}-\hat{p}_{cs2}=\frac{1}{12}\tilde{\alpha}^3\vec{x}_1.(\vec{x}_2{\times}\vec{x}_3).\nonumber\\
\end{eqnarray}
The full effective action in terms of the vectors $\vec{x}_a$ takes the form 
\begin{eqnarray}
S[\vec{x}_a]&=&\frac{c_2\tilde{\alpha}^4}{3}\bigg[\frac{1}{2}(\vec{x}_1{\times}\vec{x}_2)^2+\frac{1}{2}(\vec{x}_1{\times}\vec{x}_3)^2+\frac{1}{2}(\vec{x}_2{\times}\vec{x}_3)^2-2\vec{x}_1.(\vec{x}_2{\times}\vec{x}_3)-m^2\vec{x}_1^2-m^2\vec{x}_2^2-m^2\vec{x}_3^2\nonumber\\
&+&\frac{m^2}{2}d_{abcd}(x_1^ax_1^b+x_2^ax_2^b+x_3^ax_3^b)(x_1^cx_1^d+x_2^cx_2^d+x_3^cx_3^d)\bigg]+\bar{S}[\vec{x}_a].
\end{eqnarray}
$\bar{S}[\vec{x}_a]$ is the quantum action obtained by integrating out the field $\bar{X}_a$ ( equation (\ref{xbar}) ) from the theory. The 
coefficients $d_{abcd}$ can be computed easily from the definition $\frac{1}{N}TrL_aL_bL_cL_d=\frac{c_2^2}{3}d_{abcd}$. The original $U(N)$ gauge symmetries are now implemented by $O(3)$ orthogonal symmetries which take the $3-$dimensional vectors $\vec{x}_a$ to $\vec{x}_a^R=R\vec{x}_a$. The effect of quantum fluctuations in this problem is to  deform the shape of the parallelepiped. In particular the first order phase transition from the fuzzy sphere to the matrix phase is now seen as the transition where the parallelepiped collapses. In terms of the lengths squared $l^2$, the areas of the faces squared $a^2$ and the volume $v$ we have
\begin{eqnarray}
l^2=\frac{1}{c_2}<TrX_1^2>=\left\{\begin{array}{cc}
         \frac{\tilde{\alpha}^2}{3}& {\rm fuzzy~sphere~phase}  \\
    l^2_m & {\rm matrix~ phase}.\end{array}\right\},{\rm etc}.
\end{eqnarray}

 \begin{eqnarray}
a^2=-\frac{N}{c_2}<Tr[X_1,X_2]^2>=\left\{\begin{array}{cc}
         \frac{\tilde{\alpha}^4}{3}& {\rm fuzzy~sphere~phase}  \\
    a^2_m & {\rm matrix~ phase}.\end{array}\right\},{\rm etc}.
\end{eqnarray}

 \begin{eqnarray}
v=-\frac{i\sqrt{N}}{4c_2}<TrX_1[X_2,X_3]>=\left\{\begin{array}{cc}
         \frac{\tilde{\alpha}^3}{12}& {\rm fuzzy~sphere~phase}  \\
    v_m & {\rm matrix~ phase}.\end{array}\right\}.
\end{eqnarray}
From the data we can see that the areas of the faces squared $a^2_m$ and the volume $v_m$ in the matrix phase are constant approaching the values $2$ and $0$ respectively for small values of the coupling constant $\tilde{\alpha}$.  However the length  squared $l^2_m$ scales as $N^{-\frac{3}{2}}$ and thus it becomes $0$ in the limit. 
\subsection{Probability distribution}

As we said before for $m=0$ we can take $X_a$ to be traceless without any loss of generality and consider only the probability distribution and the partition function given by
\begin{eqnarray}
{\cal P}[X_a]=\frac{\delta\big(TrX_a\big)e^{-S[X_a]}}{Z[0]}~,~Z[0]=\int [dX_a] {\delta}\big(TrX_a\big)e^{-S[X_a]}.
\end{eqnarray}
The classical absolute minimum of the model is given by $X_a=\alpha L_a$. The quantum minimum is given by $X_a=\alpha {\phi}L_a$ where $\alpha \phi$ plays the role of the radius of the sphere with a classical value equal $\alpha$.The complete one-loop effective potential in this configuration is given in the large $N$ limit by the  formula (\ref{formula}).
The solution $\phi$ of the  equation of motion ${\phi}^4-{\phi}^3+\frac{2}{\tilde{\alpha}^4}=0$ approaches the classsical value $1$ as one increases the coupling constant $\tilde{\alpha}$ much above the critical value $\tilde{\alpha}_{*}$. Indeed it is not difficult to check that up to the order of $\frac{1}{\tilde{\alpha}^{8}}$ we have
\begin{eqnarray}
\phi=1-\frac{2}{\tilde{\alpha}^4}-\frac{12}{\tilde{\alpha}^8}+O\big(\frac{1}{\tilde{\alpha}^{12}}\big).\label{con}
\end{eqnarray} 
In this section we report on the  measurement of the radius of the fuzzy sphere. A natural definition of the radius of ${\bf S}^2_N$ is given by the observable 
\begin{eqnarray}
R^2=\frac{1}{{\alpha}^2c_2}<\frac{1}{N}\sum_{a}TrX_a^2>.
\end{eqnarray} 
The aim now is to make a precise measurement of ${\phi}$ by measuring $R^2$ and its probability distribution ${\cal P}(R^2)$.  Numerically we thermalize and then we take $Tmont$ measurements of $R^2$, we determine the minimum and maximum values $R^2_{mi}$ and $R^2_{ma}$ respectively and divide the interval $[R^2_{mi},R^2_{ma}]$ into $q=2^6+1$ smaller intervals of equal length $\delta=\frac{R^2_{ma}-R^2_{mi}}{q}$. For every measurement $R^2_{i}$,$i=1,...,Tmont$, we compute the integer
\begin{eqnarray}
j=\bigg {|}{\rm integer~part}\bigg(\frac{R^2_{i}-R^2_{mi}}{\delta}\bigg)\bigg{|}.
\end{eqnarray} 
It is clear that the value $R^2_{i}$ will lie exactly in the $j-$th interval, in other words
\begin{eqnarray}
R^2_{i}=R^2_{mi}+j{\delta}.
\end{eqnarray}
We count the number of times $N(j)$ we get the value $R^2_{i}$ and we define the corresponding probability ${\cal P}(j)$ by
\begin{eqnarray}
{\cal P}(j)=\frac{N(j)}{Tmont}.\label{prob}
\end{eqnarray}
Remark that this probability satisfies $\sum_{j=0}^q{\cal P}(j)=1$. In other words for all $j=0,...,q$ we have  ${\cal P}(j){\leq}1$.

We observe two radically different behaviour depending on wether we are inside the fuzzy sphere phase or the matrix phase. Figure $16$  shows the probability distribution in the fuzzy sphere phase whereas figure $17$ shows the probability distribution in the matrix phase.

Once again we find a good agreement between the theory and the simulation in the fuzzy sphere phase. More precisely we find that the value of $R^2$ depends explicitly but slowly on the coupling constant $\tilde{\alpha}$ but it does not depend on $N$.  In figure $13$ we see clearly that the value  of $R^2$ at the peak of the probability ${\cal P}$ is increasing with increasing $\tilde{\alpha}$. We also observe clearly how the value of $R^2$ at the peak is more or less the same for a given value of $\tilde{\alpha}$ with different $N$. These results are consistent with equation (\ref{con}). 

In figure $17$ we plot the probability distribution ${\cal P}$ as a function of $r^2={\alpha}^2c_2R^2$ for $N=6,8$ and for values of the coupling constant which are less than $2$. In other words we are inside the matrix phase. For $\tilde{\alpha}=0.5$ , $1$ and $1.5$ and  for all the values of $N$ we observe that the probability distribution in this phase peaks essentially around the same value which is estimated to be in the range $r^2=2.4-2.8$. Hence we can conclude immediately that for a fixed value of the coupling constant $\alpha$ inside the matrix phase  the order parameter $R^2=\frac{r^2}{{\alpha}^2c_2}$ will be peaked around smaller and smaller values as we increase $N$.  This means in particular that the Chern-Simons-like term in the action is playing no role in this matrix phase and as a consequence  we have no an underlying spacetime structure of a fuzzy sphere. 

The results are summarized as follows:

 \begin{eqnarray}
R^2=\frac{1}{{\alpha}^2 c_2}<\frac{1}{N}\sum_{a=1}^3TrX_a^2>=\left\{\begin{array}{cc}
         1& {\rm fuzzy~sphere~phase}  \\
    0 & {\rm matrix~ phase}.\end{array}\right\}.
\end{eqnarray}

\begin{figure}[h]
\begin{center}
\includegraphics[width=7cm,angle=-90]{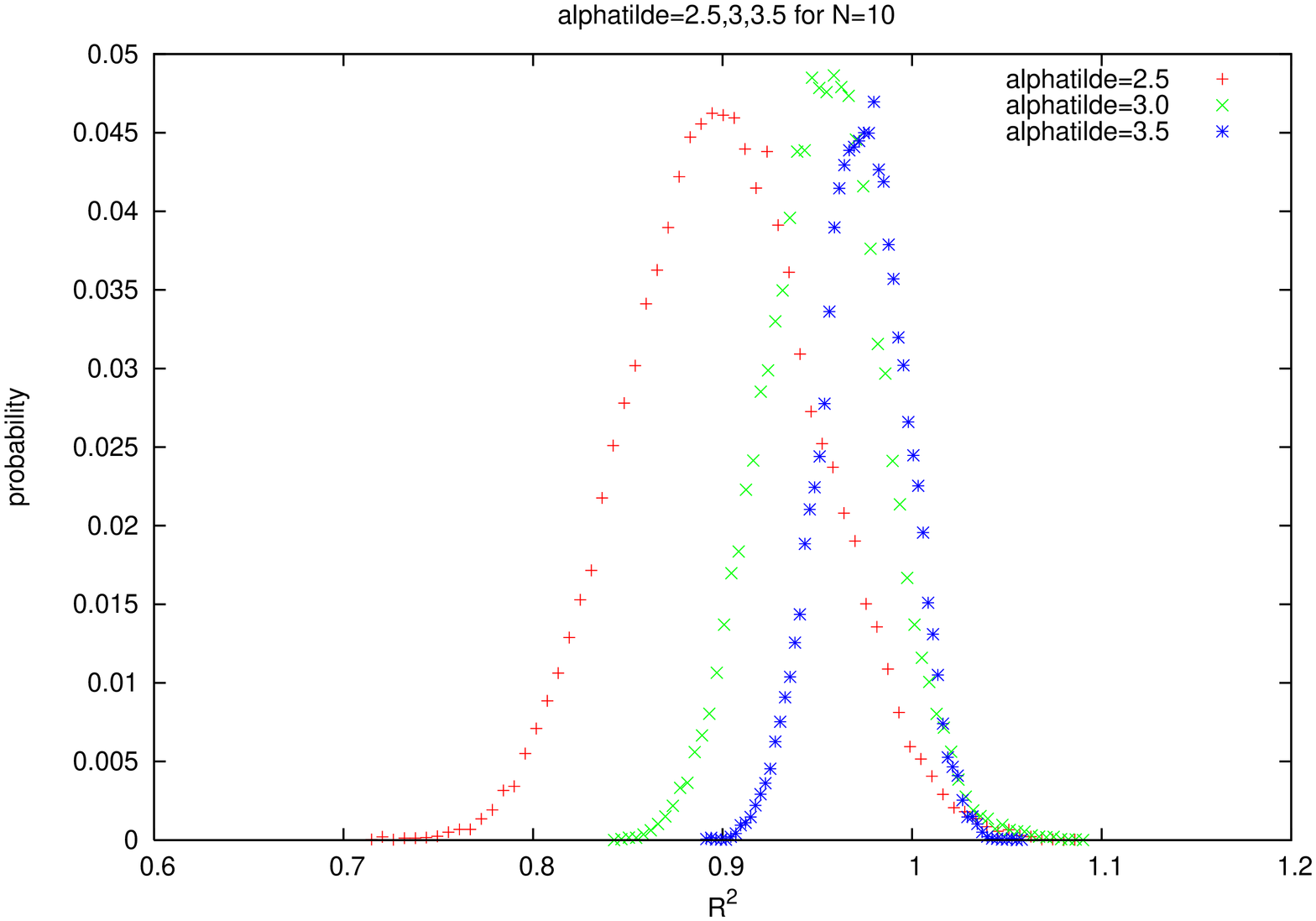}
\includegraphics[width=7cm,angle=-90]{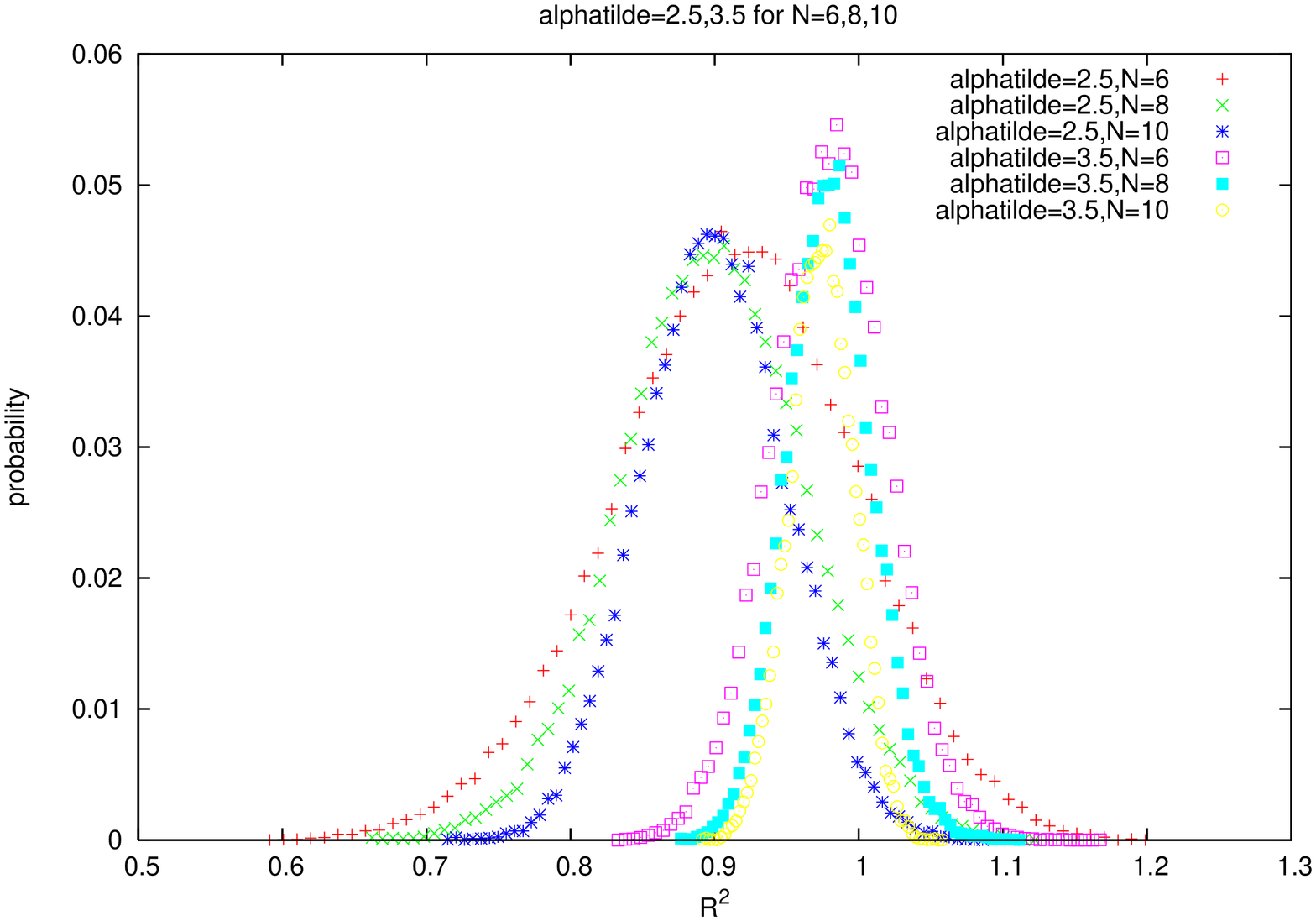}
\caption{{ The probability distribution in the fuzzy sphere phase.}}
\end{center}
\end{figure}

\begin{figure}[h]
\begin{center}
\includegraphics[width=7cm,angle=-90]{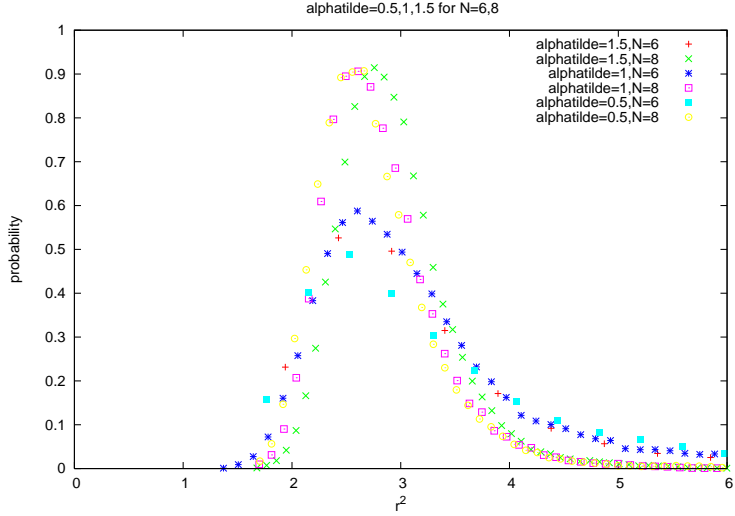}
\caption{{ The probability distribution in the matrix phase.}}
\end{center}
\end{figure}

\bibliographystyle{unsrt}

\end{document}